\documentstyle[12pt]{article}
\newcommand{\be}{\begin{equation}}
\newcommand{\ee}{\end{equation}}
\newcommand{\tepro}{{\bf \otimes}}
\newtheorem{definition}{Definition}
\newtheorem{theo}{Theorem}
\newtheorem{lemma}{Lemma}
\newtheorem{example}{Example}
\begin{document}
\title{An introduction to quantized Lie groups and
algebras}
\author{T.Tjin \\ Instituut voor Theoretische Fysica  \\
Valckenierstraat 65 \\ 1018 XE Amsterdam \\ The Netherlands}
\date{November 1991}
\maketitle
\begin{abstract}
We give a selfcontained introduction to the
theory of quantum groups according to Drinfeld highlighting
the formal aspects as well as the applications to the
Yang-Baxter equation and representation theory.
Introductions to Hopf algebras, Poisson structures and
deformation quantization are also provided.
After having defined Poisson-Lie groups we study their relation
to Lie-bi algebras and the classical Yang-Baxter equation. Then
we explain in detail the concept of quantization for them. As
an example the quantization of $sl_2$ is explicitly carried out.
Next we show how quantum groups are related to the Yang-Baxter
equation and how they can be used to solve it. Using the
quantum double construction we explicitly construct the
universal $R$-matrix for the quantum $sl_2$ algebra.
In the last section we deduce all finite dimensional irreducible
representations for $q$ a root of unity. We also give their tensor
product  decomposition (fusion rules) which is relevant to
conformal field theory.
\end{abstract}

\section*{Introduction}
In the beginning of the eighties
important progress was being made in the field
of quantum integrable field theories. Crucial to this
was a quantum mechanical version of the well known inverse
scattering method used so successfully in the theory of
integrable non-linear evolution equations like the Korteweg
de Vries equation (KdV). In \cite{KuRe} P.Kulish and Y.Reshetikhin
showed that the quantum linear problem of the quantum
sine-Gordon equation was not associated with the Lie algebra
$sl_2$ as in the classical case, but with a deformation of this
algebra. Other work showed \cite{Skl} that deformations
of Lie algebraic structures were not special to the quantum
sine-Gordon equation and it seemed that they were part of a
more general theory.

It was V.I.Drinfeld who showed that  a suitable algebraic
quantization of so called Poisson Lie groups reproduced exactly
the deformed algebraic structures encountered in the theory of
quantum inverse scattering \cite{Dr1,Dr2,Dr3}.
At approximately the same time
M.Jimbo derived the same relations coming from a slightly
different direction \cite{Jim1,Jim2}.
The new algebraic structures
were called quantized universal enveloping algebras (QUEA) and
have been the object of intense investigation by mathematicians
as well as physicists. The main reason that $QUEA$ are of such
great importance is that they are closely related to the so called
quantum Yang-Baxter equation which plays a prominent role in many
areas of research such as knot theory, solvable lattice models,
conformal field theory and quantum integrable systems. There was
great excitement when string theorists found out that the decomposition
of tensor product representations of the QUEAs resembled very much
 the fusion rules of certain conformal field theories. This
triggered a great deal of research on the relation between
quantum groups and chiral conformal algebras \cite{Conf}.
Unfortunately
the relation turned out to be not  as straightforward as hoped
and even today there is no completely
satisfying answer to the question how precisely
conformal field theory is related to quantum groups. At the
moment research focusses on the representation theory
of QUEAs.

A different approach to quantum groups and one with a completely
different background (C$^{\ast}$-algebra theory) was initiated
by S.L.Woronowicz who initially called them pseudogroups
\cite{Wor1,Wor2,Wor3,Man}.
The crucial ingredient in this approach is the Gelfand-Naimark
theorem which roughly states that any commutative $C^{\ast}$-algebra
with unit element is isomorphic to an algebra of all continuous
functions on some compact topological manifold. If the topological
space is a topological group then the space of functions on it
picks up extra structure. Generalizing the basic properties of
function spaces on topological groups to non-commutative
$C^{\ast}$-algebras the interpretation in terms of an underlying
manifold is gone but we can still persue the theory. The
motivation for this is that in the commutative case all algebraic
information on the topological group is contained in the (extra)
structure of its function space which means that we can study
the group manifold itself or its function space, it does not make
any difference. In the non-commutative case we only have information
on the function space (if you insist on calling it that), but
this only means that you no longer have the same information in
two different disguises. This approach to quantum groups is the
one which is most popular among pure mathematicians (algebraists).

Even though the two approaches have different origins they are
closely related \cite{Bur1}.
The main differences are that in the
Drinfeld approach one does not quantize the space of functions
on a group (at least not directly) but the universal enveloping
algebra of the Lie algebra which is dual to the space of functions.
However a more important difference is that in the
Woronowicz approach there is no mention of a
Poisson structure while they play a prominent role in the
Drinfeld approach.

Up to now Drinfeld's approach to quantum groups has
received the most attention in the physics literature.
The
present paper is meant to give a non-specialist
introduction to this approach and its
applications providing proofs and derivations where
they are omitted in literature. In section 1
we review the basic facts of Hopf algebras which is the language
in which the theory of quantum groups is written.
We also show that the space of functions on a Lie group is a
commutative Hopf algebra. As mentioned above it contains all
essential information  on $G$, and its dual space is shown to be
'almost equal' to the universal enveloping algebra of the Lie algebra
of $G$, which is also a Hopf algebra. We thus go from the group $G$
, via the space of smooth functions on $G$ to the
universal enveloping algebra of the Lie algebra of $G$.
The reason for taking this path becomes clear in sections 2 and 3
where we introduce and discuss in detail Poisson and co-Poisson
structures. If the Lie group happens to be the phase space of some
classical dynamical system, then the space of functions on it
carries a Poisson bracket. This in turn induces a co-Poisson bracket
on the universal enveloping algebra because of the duality between
these two spaces. In short, the universal enveloping algebra of a
Lie group which is also a classical phase space is a co-Poisson
Hopf algebra (section 3). In section 4 there is then defined a suitable
form of quantization for such an object which involves, as is well
known in physics, replacing a (co)Poisson bracket by a
(co)commutator. Since the first four sections are rather abstract we
consider in section 5 the case of $sl_2$ as an example. Using the
definition of quantization given in section 4 we quantize $sl_2$
explicitly giving the so called q-deformed algebra ${\cal U}_q(sl_2)$.
In section 6 we use this algebra to explicitly solve the Yang-Baxter
equation which is of interest to many areas of physics.
The solution $R$ we thus obtain is a 'universal object' i.e. it
is ${\cal U}_{q}(sl_2) \tepro {\cal U}_{q}(sl_2)$ valued. In order
to make it more useful to physical applications we need to
consider representations of ${\cal U}_q (sl_2)$ which then turn $R$
into a matrix (a so called $R$-matrix). This is the subject of the
last section where we consider the representation theory of
${\cal U}_q(sl_2)$ at '$q$ a root of unity'. We only consider
the representation theory in this regime because it seems the
most interesting from a physical point of view.  We
deduce all finite dimensional irreducible representations of
${\cal U}_q(sl_2)$ and also give the structure of their
tensor product representations. We also hint on a relation between
this and the fusion rules of certain conformal field theories.

We have tried to keep the paper as selfcontained as possible.
The reader is assumed to know something about Lie algebras and
differential geometry.

\section{Hopf algebras}
Any selfcontained review paper on quantum groups is bound to
start with the definition and elementary properties of
Hopf algebras \cite{Abe,Mil}.
The reason for this is that, as we will see
in detail, the algebra of functions on a Lie group (which is
the object of any quantization attempt) is a commutative
Hopf algebra. As we already mentioned in the introduction,
this Hopf algebra contains all the information on the algebraic
structure of the Lie group. Therefore quantizing the Lie group as
a manifold and as an algebraic structure means deforming the
Hopf algebra structure of the function space while maintaining
the fact that it is a Hopf algebra.

Instead of just giving the definition of a Hopf algebra,
 which is rather involved, we will introduce its structure step
by
step.
First let us give a definition of an algebra.
\begin{definition}
An algebra is a linear space $A$ together with two maps
\begin{eqnarray}
m & : &A {\bf \otimes } A \rightarrow A \\
\eta & : &{{\bf C}}
\rightarrow A
\end{eqnarray}
such
that
\begin{enumerate}
\item $m$ and $\eta$ are linear
\item $m(m {\bf \otimes } 1)=m(1 {\bf \otimes} m)$ (associativity)
\item $m(1 {\bf \otimes} \eta)=m(\eta {\bf \otimes} 1)=id$ (unit)
\end{enumerate}

\end{definition}
We will usually denote $m(a {\bf \otimes }b)$ by $a.b$.

Properties 2 and 3 imply that the diagrams in figure 1 commute.
\begin{figure}
\centering
\begin{picture}(250, 400)
\put(10, 110){\makebox(0, 0)[r]{${{\bf C}}\;\bigotimes A$}}
\put(240, 110){\makebox(0, 0)[l]{$A\;\bigotimes{{\bf C}}$}}
\put(125, 110){\makebox(0, 0){$A\;\bigotimes A$}}
\put(125, 20){\makebox(0, 0){$A$}}
\put(20, 110){\thicklines\vector(1, 0){75}}
\put(230, 110){\thicklines\vector(-1, 0){75}}
\put(55, 123){\makebox(0, 0){\normalsize$\eta\;\bigotimes1$}}
\put(185, 123){\makebox(0, 0){\normalsize$1\;\bigotimes\eta$}}
\put(125, 90){\thicklines\vector(0, -1){50}}
\put(130, 65){\makebox(0, 0)[l]{\normalsize$m$}}
\put(10, 100){\thicklines\line(6, -5){75}}
\put(240, 100){\thicklines\line(-6, -5){75}}
\put(30, 65){\makebox(0, 0){\Huge$\sim$}}
\put(215, 65){\makebox(0, 0){\Huge$\sim$}}

\put(10, 310){\makebox(0, 0){$A\;\bigotimes A\;\bigotimes A$}}
\put(50, 310){\thicklines\vector(1, 0){140}}
\put(120, 323){\makebox(0, 0){\normalsize$m\bigotimes1$}}
\put(210, 310){\makebox(0, 0){$A\;\bigotimes A$}}
\put(10, 300){\thicklines\vector(0, -1){70}}
\put(5, 265){\makebox(0, 0)[r]{\normalsize$1\bigotimes m$}}
\put(210, 300){\thicklines\vector(0, -1){70}}
\put(215, 265){\makebox(0, 0)[l]{\normalsize$m$}}
\put(10, 220){\makebox(0, 0){$A\;\bigotimes A$}}
\put(50, 220){\thicklines\vector(1, 0){140}}
\put(120, 233){\makebox(0, 0){\normalsize$m$}}
\put(210, 220){\makebox(0, 0){$A$}}
\end{picture}  \caption{The commutativity of these diagrams
expresses the algebra axioms}
\end{figure}
(here $\sim$ refers to the fact that ${{\bf C}}\tepro A
\equiv {{\bf C}}\tepro_{{{\bf C}}}A \simeq A$ by the obvious
isomorphism $\lambda \tepro a = \lambda a$ (for $\lambda
\epsilon {{\bf C}}$, and $a \epsilon A$)).
Property 3 is an unusual way of saying that $A$ has a unit element,
for let $\alpha \tepro a \epsilon A \tepro {\bf C} \simeq A$, then
$m(\eta \tepro 1 ) (\alpha \tepro a ) = \eta (\alpha ) a$ which
by property 3 is equal to $\alpha a$. This means that $\eta (\alpha )
= \alpha . 1$ where 1 is the unit element of $A$.

A precise name for the algebras
defined above would be unital associative algebras
, however we will simply call them algebras.
It is easy to think of all sorts of examples (group algebras,
rings, fields etc.).

Let $(A, m_A, \eta_A)$ and $(B, m_B, \eta_B)$ be algebras, then
the tensor product space $A \tepro B$ is naturally endowed
with the structure of an algebra. The multiplication
$m_{A \tepro B}$ on $A \tepro B$ is defined by
\be
m_{A \tepro B}=(m_A \tepro m_B )( 1\tepro \tau \tepro 1)
\ee
where $\tau$ is the so called flip map $\tau (a \tepro b)=
b \tepro a$. More explicitly this multiplication on
$A \tepro B$ reads $(a_1 \tepro b_1).(a_2 \tepro b_2)=
(a_1 . a_2) \tepro (b_1 . b_2)$. It follows that the set of
algebras is closed under taking tensor products.

Even though these abstract algebras are of considerable importance
themselves,  people (that is physicists) are primarily interested in
their representation
theory. By a representation we mean the usual, i.e.
a homomorphism from the algebra $A$
to an algebra of linear operators on
some vectorspace. For completeness we give the definition.
\begin{definition}
Let $(A, m, \eta )$ be an algebra, $V$ a linear space and $\rho$ a map
from $A$ to the space of linear operators in $V$. $(V, \rho)$ is
called a representation of $A$
if
\begin{enumerate}
\item $\rho$ is linear
\item $\rho (xy)= \rho (x) \rho (y)$
\end{enumerate}

\end{definition}
In physics it often happens that one has to compose two
representations (for example when adding angular momenta).
This happens when two physical systems each within a certain
representation interact (for example two spin $1/2$ particles).
Mathematically this means that one must consider tensorproduct
representations of the underlying abstract algebra. Let us now
try to define tensor product representations for the algebras
defined above.

Suppose $(\psi_1, V_1)$ and$(\psi_2, V_2)$ are two representations
of an algebra $A$. How do we define an action of $A$ an
$V_1 {\bf \otimes}V_2$ using $\psi_1$ and $\psi_2$ ? There
are only two reasonable
possibilities:
\begin{enumerate}
\item The action of $a\epsilon A$ on
$v_1 {\bf \otimes} v_2 \epsilon V_1 {\bf \otimes }V_2$ is:
\be
a.(v_1 {\bf \otimes }v_2 )=(\psi_1(a)v_1) {\bf \otimes}
(\psi_2(a)v_2)
\ee
\item The action of $a$ on $v_1 {\bf \otimes }v_2$ is
\be
a.(v_1 {\bf \otimes }v_2)=\psi_1(a)v_1 {\bf \otimes}v_2 +
v_1 {\bf \otimes }\psi_2 (a)v_2
\ee
\end{enumerate}

Definition
1 certainly does not satisfy the required properties since
such a tensor product representation would not be linear.
Definition 2 has the problem that the homomorphism
property does not hold unless the multiplication on $A$ is
antisymmetric (as is the case in for example a Lie algebra).
There seems no way out, we have to endow the algebra with
some extra structure which makes it possible to define tensor
product representations.

Consider a map $\Delta:A\rightarrow A {\bf \otimes}A$ and define
the tensor product representation $\Psi$ by
\be
\Psi=(\psi_1 {\bf \otimes}\psi_2)\Delta
\ee
We require of $\Psi$ that it be linear, satisfy the
homomorphism property and also that the representations
$(V_1 {\bf \otimes}V_2) {\bf \otimes }V_3$ and
$V_1 {\bf \otimes }(V_2 {\bf \otimes }V_3)$ are equal
(in this way the set of representations becomes a ring). These
requirements lead to the following conditions on $\Delta$
:
\begin{enumerate}
\item $\Delta$ is linear.
\item $\Delta (ab)=\Delta (a) \Delta (b)$
\item $(\Delta {\bf \otimes }id)\Delta=
      (id {\bf \otimes }\Delta)\Delta$
\end{enumerate}

A map $\Delta : A \rightarrow A \tepro A$ with these properties
is called a co-multiplication. Property 3 means that the
diagram in figure 2 commutes.
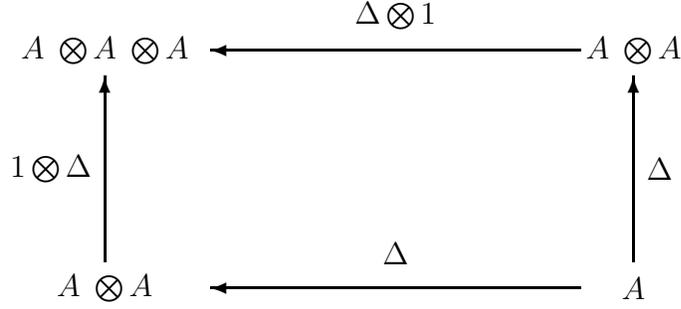
\begin{figure}
\centering
\begin{picture}(250, 125)
\put(10, 110){\makebox(0, 0){$A\;\bigotimes A\;\bigotimes A$}}
\put(190, 110){\thicklines\vector(-1, 0){140}}
\put(120, 123){\makebox(0, 0){\normalsize$\Delta\bigotimes1$}}
\put(210, 110){\makebox(0, 0){$A\;\bigotimes A$}}
\put(10, 30){\thicklines\vector(0, 1){70}}
\put(5, 65){\makebox(0, 0)[r]{\normalsize$1\bigotimes\Delta$}}
\put(210, 30){\thicklines\vector(0, 1){70}}
\put(215, 65){\makebox(0, 0)[l]{\normalsize$\Delta$}}
\put(10, 20){\makebox(0, 0){$A\;\bigotimes A$}}
\put(190, 20){\thicklines\vector(-1, 0){140}}
\put(120, 33){\makebox(0, 0){\normalsize$\Delta$}}
\put(210, 20){\makebox(0, 0){$A$}}
\end{picture} \caption{Co-associativity}
\end{figure}
Compare this diagram to the one which expressed the
associativity of the multiplication map. The only difference
is that all the arrows are reversed. Property 3 is therefore
called co-associativity. Property 2 states that $\Delta$
is an algebra homomorphism.

Associated with the multiplication on $A$ we had a map $\eta$
whose properties signaled the existence of a unit in $A$.
In the same way we now define a map, called a co-unit, which
has properties that follow by reversing all arrows in the
diagram for the map $\eta$ (see figure 3)

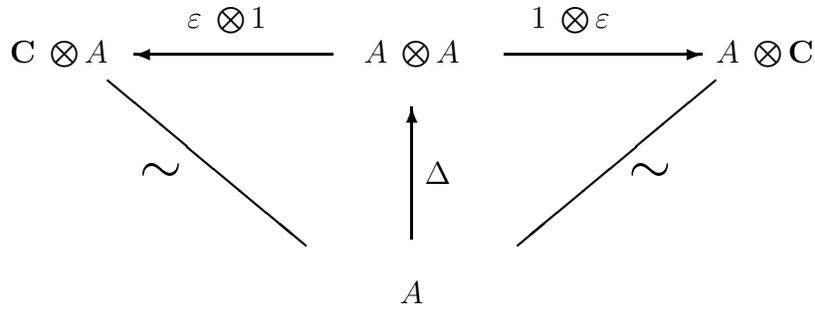
\begin{figure}
\centering
\begin{picture}(250, 150)

\put(10, 110){\makebox(0, 0)[r]{${{\bf C}}\;\bigotimes A$}}
\put(240, 110){\makebox(0, 0)[l]{$A\;\bigotimes{{\bf C}}$}}
\put(125, 110){\makebox(0, 0){$A\;\bigotimes A$}}
\put(125, 20){\makebox(0, 0){$A$}}
\put(95, 110){\thicklines\vector(-1, 0){75}}
\put(160, 110){\thicklines\vector(1, 0){75}}
\put(55, 123){\makebox(0, 0){\normalsize$\varepsilon\;\bigotimes1$}}
\put(185, 123){\makebox(0, 0){\normalsize$1\;\bigotimes\varepsilon$}}
\put(125, 40){\thicklines\vector(0, 1){50}}
\put(130, 65){\makebox(0, 0)[l]{\normalsize$\Delta$}}
\put(10, 100){\thicklines\line(6, -5){75}}
\put(240, 100){\thicklines\line(-6, -5){75}}
\put(30, 65){\makebox(0, 0){\Huge$\sim$}}
\put(215, 65){\makebox(0, 0){\Huge$\sim$}}

\end{picture} \caption{The co-unit}
\end{figure}

\begin{definition}
A co-unit is a map $\varepsilon :A\rightarrow {{\bf C}}$ such that
$(1 {\bf \otimes}\varepsilon)\Delta = (\varepsilon {\bf \otimes}1
)\Delta =id $
\end{definition}

The set $(A, m, \Delta, \eta, \varepsilon)$ is called a
bi-algebra if all the above properties are satisfied.
It follows from this definition that we know how to define
tensor product representations for bi-algebras.

As we will see in the examples certain bi-algebras are
related to Lie-groups. It turns out that we can encode
all the properties of Lie groups into a particular class
of bi-algebras. First of all these bi-algebras are commutative
(i.e. $a.b=b.a$). Also they possess an extra structure called
an antipode or co-inverse which is (the name already gives it
away) related to the fact that every element of a group has an
inverse (or put more formally, there exists a map $I$ from the
group to itself such that $I(g)g=gI(g)=e$ where $e$ is the
unit element of the group). Considering bi-algebras
associated to Lie groups as a special case, the properties
of the antipode can be generalized. This leads to the following
\begin{definition}
A Hopf algebra is a bi-algebra $(A, m, \eta, \Delta, \varepsilon)$
together with a map
\be
S:A \rightarrow A
\ee
with
the following property:
\be
m(S {\bf \otimes }id)\Delta=m(id {\bf \otimes} S)\Delta
=\eta \circ \varepsilon
\ee
$S$ is called an antipode.
\end{definition}

In order to show the relevance of Hopf algebras to the theory
of quantum groups we will now consider two examples which
are crucial to the understanding of this theory.

\begin{example}
Let $G$ be a compact topological group. Consider the space of
continuous functions on $G$ denoted by $C(G)$ together with
the following maps:
\begin{itemize}
\item $(f.h)(g)=f(g)h(g)$
\item $\Delta (f)(g_1 {\bf \otimes }g_2)=f(g_1 g_2)
$
\item $\eta (x) =x1$ where 1(g)=1 for all $g \epsilon G
$
\item $\varepsilon (f)=f(e)$ where $e$ is the unit element of
$G$
\item $S(f)(g)=f(g^{-1})$
\end{itemize}
where $g_1, g_2, g \epsilon G$, $x\epsilon {{\bf C}}$ and
$f, h \epsilon C(G)$.
It is easy to check that
the set $C(G)$ together with these maps is a Hopf algebra.
Moreover note that it is a commutative Hopf algebra. Therefore
we have associated a commutative Hopf algebra to every
compact Topological group. You might wonder if is possible
conversely to associate a topological group to a given
commutative Hopf algebra such that the group which
is thus associated to the Hopf algebra $C(G)$ is the
group $G$. As mentioned in the introduction
the answer to this question is contained in a
theorem by Gelfand and Naimark and is affirmative (actually
they proved their theorem within the context of the
$C^{\ast}$-algebras so it is necessary that we have given
on the Hopf algebra also a $\ast$-involution and
a norm that turn the Hopf algebra into
a $C^{\ast}$ algebra).
The group is constructed as the set of
characters of the $C^{\ast}$-Hopf algebra.
We will not go into this
however.

Doing quantum-groups the non-commutative geometric way means
deforming (i.e. making non-commutative) in some way the Hopf
algebra of functions on the group. This is studied in detail
in \cite{Wor1,Wor2,Wor3}.
\end{example}
As became clear in
the above example the set of commutative
Hopf algebras is a very important subset. Another important
subset, and in some way a dual one, is the subset of
universal enveloping algebras. This is the example we will
take a look at next.
\begin{example}
Let $L$ be a Lie algebra and ${\cal U}(L)$ its universal
enveloping algebra, then ${\cal U}(L)$ becomes a Hopf
algebra if we define
\begin{itemize}
\item The multiplication is the ordinary multiplication in
${\cal U}(L)$.
\item $\Delta (x)=x \tepro 1 + 1\tepro x$
\item $\eta (\alpha)= \alpha 1$
\item $\varepsilon (1)=1$ and zero on all other elements.
\item $S(x)=-x$
\end{itemize}
where $x$ is an element of $L$ (considered as a subset of
${\cal U}(L)$). Strictly speaking this defines $\Delta, \eta,
\varepsilon$ and $S$ only on the subset $L$ of the universal
enveloping algebra, however it is easily seen that these
maps can be extended uniquely to all of ${\cal U}(L)$ such
that the Hopf algebra axioms are satisfied everywhere. Note
that an arbitrary Lie algebra
 $L$ itself
is not a Hopf algebra because it is not an
associative algebra. Also note that the universal enveloping
algebras are co-commutative Hopf algebras, that is we have
the equality $\Delta=\tau \circ \Delta$ where $\tau$ is again the
flip operator.
\end{example}
The two examples given above are in a sense dual. In order to
explain what we mean by this  we will now
show that the dual space of a Hopf algebra is again a Hopf
algebra. Suppose that $(A, m, \Delta, \eta, \varepsilon, S)$ is a
Hopf algebra and that $A^{\ast}$ is its dual space, then using
the structure maps of $A$ we  define the structure maps
$(m^{\ast}, \Delta^{\ast}, \eta^{\ast}, \varepsilon^{\ast}S^{\ast}
)$ on
$A^{\ast}$ as follows:
\begin{itemize}
\item $\langle m^{\ast}(f \tepro g), x \rangle =
\langle f \tepro g , \Delta (x)\rangle$
\item $\langle \Delta^{\ast}(f), x \tepro y\rangle=
\langle f, xy \rangle $
\item $\langle \eta^{\ast}(\alpha), x\rangle = \alpha .
\varepsilon (x) $
\item  $\varepsilon^{\ast}(f)=\langle f, 1 \rangle $
\item $\langle S^{\ast}(f), x \rangle =
\langle f, S(x)\rangle$
\end{itemize}
where $f$ and $g$ are elements of $A^{\ast}$ and $x, y$ are
elements of $A$. The brackets $\langle ., . \rangle$ denote
the dual contraction between $A^{\ast}$ and $A$, and
$\langle f \tepro g, x \tepro y \rangle =\langle f, x\rangle
\langle g, y\rangle$. It is easy, using the fact that $A$ is a
Hopf algebra, to verify that $A^{\ast}$ is also a Hopf
algebra. Note that if $A$ is commutative then $A^{\ast}$ is
co-commutative, and if $A$ is co-commutative then $A^{\ast}$
is commutative. This is so because the multiplication on
$A$ induces the comultiplication on $A^{\ast}$, and the
comultiplication on $A$ induces the multiplication on $A^{\ast}$.

We will now discuss the duality between ${\cal U}(L)$ and
$C^{\infty}(G)$. Consider the map
\be
\rho :L \rightarrow End_{{\bf C}}(C^{\infty}(G))
\ee
defined by
\be
(\rho (X)\phi)(g)=\frac{d}{dt}(\phi ( e^{tX}g))\mid_{
t=0}
\ee
for $X \epsilon L$, $\phi \epsilon C^{\infty}(G)$, $g\epsilon G$
(the derivative in $g$ of $\phi$ in the direction $X$).
This map extends uniquely to a homomorpism
\be
\rho :{\cal U}(L) \rightarrow End_{{\bf C}}(C^{\infty}(G))
\ee
(by definition of ${\cal U}(L)$). Also consider the right action
of $G$ on $C^{\infty}(G)$
\be
R_g :C^{\infty}(G) \rightarrow C^{\infty}(G)
\ee
defined by $R_g(\phi)(g^{\prime})=\phi (g^{\prime} g)$. We have
the following lemma:
\begin{lemma}
The  map $\rho$ has the following properties.
\begin{enumerate}
\item $R_g \circ \rho (a) = \rho (a) \circ R_g \;\;\;\; $ (right
invariance)
\item $\rho (X)(\phi . \psi )=(\rho (X) \phi ).\psi +\phi .
(\rho (X)\psi ) \;\;\;\; $ (derivation property)
\end{enumerate}
for all $a \epsilon {\cal U}(L)$, $X \epsilon L$ and
$\phi, \psi \epsilon C^{\infty}(G)$.
\end{lemma}
Proof: A straightforward calculation gives
\be
(\rho (X) \circ R_g)(\phi)(g^{\prime})=
\frac{d}{dt} \phi (e^{tX}g^{\prime} g)\mid_{t=0} =
(R_g \circ \rho (X))(\phi )(g^{\prime})
\ee
for $X \epsilon L$. Using the fact that $\rho$ is an algebra
homomorphism part 1 of the lemma follows.
Part 2 follows immediately from the Leibniz rule.
This concludes the proof.

Define the pairing
\be
\langle ., . \rangle :C^{\infty}(G) \times {\cal U}(L)
\rightarrow {{\bf C}}
\ee
by $\langle \phi, a \rangle =(\rho (a) \phi )(e)$ where $e$ is
the unit element of $G$, $a \epsilon {\cal U}(L)$ and
$\phi \epsilon C^{\infty}(G)$.

\begin{theo}
The map
\be
C^{\infty}(G) \rightarrow ({\cal U}(L))^{\ast}
\ee
defined by
\be
\phi \rightarrow \langle \phi , . \rangle
\ee
is an embedding.
\end{theo}
Proof: We have to prove that this map is injective. Suppose
$\langle \phi_1 , a \rangle = \langle \phi_2 , a \rangle$
for all $a \epsilon {\cal U}(L)$.  Then
\be
0= \langle \phi_1 -\phi_2 , a \rangle = (\rho (a) (
\phi_1 - \phi_2 )) (e)
\ee
Note that $\rho (a)$ is a differential operator of arbitrary
order. This is clear from the fact that if $X \epsilon L
\subset {\cal U}(L)$ then by the above lemma $\rho (X)$ is
a derivation which means that it is a vectorfield. A
vectorfield can be seen as a differential operator of order 1
on $C^{\infty}(G)$. An element $a$ of ${\cal U}(L)$ can in
turn be seen as a linear combination of monomials of elements
of $L$. Since $\rho$ is an algebra homomorphism it follows
that $\rho (a)$ is a differential operator of arbitrary order.
What we have therefore deduced is that any derivative of the
function $\phi_1 - \phi_2$ in the point $e \epsilon G$ is zero.
Since a smooth fuction is uniquely determined by its derivatives
in $e$ we find $\phi_1 - \phi_2 =0$. This proves the theorem.

So indeed there is a duality between ${\cal U}(L)$ and
$C^{\infty}(G)$ because $C^{\infty}(G)$ can be embedded into
$({\cal U}(L))^{\ast}$.
Using the definition of the dual Hopf algebra
given above we can endow ${\cal U}(L)$ with a Hopf algebra
structure that is induced by the one on $C^{\infty}(G)$. It is
easy to check that this is precisely the Hopf algebra structure
of example 2 so the spaces in two examples are not only
dual as spaces but also as Hopf algebras.

\section{Poisson structures}
 From a mathematical point of view a classical mechanical system
is fixed by giving a phase space, which consists of a smooth
manifold together with a closed non-degenerate 2 form (a
symplectic form), and a specific function on the manifold
which plays the role of a Hamiltonian (determining the dynamics
of the system). The symplectic form determines the Poisson
bracket. In the spirit of the previous section	we
translate all the ingredients of classical mechanical systems
into an algebraic language by passing on to the space of
smooth functions on the phase space. The reason for this is
again that this algebraic approach  leaves enough room for
generalization to non-commutative algebras (in which case
there is no longer an interpretation in terms of an underlying
manifold). Our first definition will be that of a Poisson algebra.
\begin{definition}
A Poisson algebra is a commutative algebra $(A, m, \eta)$
together with a map
\be
\{., .\}:A \times A \rightarrow A
\ee
such
that
\begin{enumerate}
\item $A$ is a Lie algebra with respect to $\{., .\}$.
\item $\{ab, c\}=a\{b, c\}+\{a, c\}b$
\end{enumerate}

\end{definition}
Obviously the space of smooth functions on a symplectic
manifold is a Poisson algebra.

For later use we shall reformulate the defining properties of
a Poisson algebra somewhat. By definition of the tensorproduct
the map $\{., .\}$ induces a map
$\gamma :A \tepro A \rightarrow A$.
We can reformulate the properties of $\{., .\}$
in terms of $\gamma$
and they read
:
\begin{enumerate}
\item $\gamma \circ \tau = -\gamma \;\;$ (anti-symmetry) \\
$\gamma (1 \tepro \gamma)(1\tepro 1\tepro 1 +
(1\tepro \tau)(\tau \tepro 1)+(\tau \tepro 1)(1 \tepro \tau))=0
\;\;\;$ (Jacobi-identity)
\item $\gamma (m \tepro 1)=m(1\tepro \gamma)(1 \tepro 1
\tepro 1 +\tau \tepro 1)$
\end{enumerate}

It is straightforward to derive these identities from the
defining properties of a Poisson algebra.

Let us now consider the concept of a Poisson algebra homomorphism.
\begin{definition}
Let $(A, m_A, \{., .\}_A)$ and $(B, m_B, \{., .\}_B)$
be Poisson algebras.
A Poisson algebra homomorphism is a linear map $f$ from $A$ to $B$
such that
\begin{itemize}
\begin{enumerate}
\item $f(a.b)=f(a)f(b)$
\item $f(\{a, b\}_A)=\{f(a), f(b)\}_B$
\end{enumerate}
\end{itemize}
where $a$ and $b$ are arbitrary elements of $A$.
\end{definition}
Again we can write the above expressions entirely in terms
of $m$ and $\gamma$. As one easily verifies the first
expression becomes
\be
f \circ m_A =m_B \circ (f \tepro f)
\ee
while the second one reads
\be
f \circ \gamma_A =\gamma_B \circ (f \tepro f)
\ee

The tensor product space $A \tepro B$ of the two Poisson
algebras inherits from its constituents a natural
Poisson structure. First of all we have to say how
to multipy two elements of $A \tepro B$. This is easy
\be
(a \tepro b).(c \tepro d)=(a.b) \tepro (c.d)
\ee
where $a, c \epsilon A$ and $b, d \epsilon B$, or
equivalently
\be
m_{A \tepro B}=(m_A \tepro m_B ) (1 \tepro \tau \tepro 1)
\ee
where $\tau$ is again the flip operator.
Second we have to define a Poisson structure such that the
axioms of a Poisson algebra are satisfied. The following
Poisson bracket does the trick
\be
\{a\tepro b, c\tepro d \}_{A \tepro B} = \{a, c\} \tepro bd +
ac \tepro \{b, d\}
\ee
or in other words
\be
\gamma_{A \tepro B}=(\gamma_A \tepro m_B +m_A \tepro \gamma_B)
(1 \tepro \tau \tepro 1) \label{TP}
\ee
So the set of Poisson algebras is closed under taking the
tensorproduct.

As we argued earlier physicists will primarily be interested in
bi-algebras because for them we know how to define tensorproduct
representations. For a Poisson algebra we do not want any
co-product however because we want a tensor
product representation (defined through the co-product)
to be a Poisson algebra homomorphism not merely an algebra
homomorphism. This is easily seen to give the following
condition on $\Delta$
\be  \label{comp}
\{\Delta (a) , \Delta (b) \}_{A \tepro A}=\Delta (\{a, b\}_A)
\ee
or equivalently
\be
\Delta_{A \tepro A} \circ (\Delta \tepro \Delta )= \Delta \circ
\gamma_A
\ee
If this is satisfied then $A$ is called a Poisson bi-algebra.
(If the algebra $A$ by accident also carries an antipode,
then you give it credit by calling $A$
a Poisson Hopf algebra.)

By now everything has become pretty algebraic and soon we will
be able to profit from this. However we still are not where
we want to be. First we have to introduce the concept of
a co-Poisson structure since we will need this in the theory
of quantum groups. As usual the 'co'
means that something gets dualized, in this case the
Poisson structure.
We have to dualize again because of the duality between
the universal enveloping algebra of the Lie group
 and the space of smooth functions on the Lie group.
Therefore a Poisson bracket on the Lie group will be a
co-Poisson structure on the universal enveloping algebra.

Here is the precise definition of a co-Poisson bi-algebra
\begin{definition}
A co-Poisson bi-algebra is a co-commutative
bi-algebra $(A, m, \Delta, \eta,
\varepsilon )$	together with a map
\be
\delta :A \rightarrow A \tepro A
\ee
such that
\begin{enumerate}
\item $\tau \circ \delta = - \delta$ (co-antisymmetry)
\item $(1 \tepro 1 \tepro 1 +(1 \tepro \tau)(\tau \tepro 1)
+(\tau \tepro 1)( 1 \tepro \tau ))(1 \tepro \delta )\delta =0$
(co-Jacobi id.)
\item $(\Delta \tepro 1)\delta =(1 \tepro 1 \tepro 1 +\tau \tepro 1
)(1 \tepro \delta ) \Delta$ (co-Leibniz rule)
\item $(m \tepro m) \circ \delta_{A \tepro A}=\delta \circ m \;\;$
(i.e. $m$ is a co-Poisson
homomorphism)
\end{enumerate}
where $\delta_{A \tepro A}=(1 \tepro \tau \tepro 1)(\delta
\tepro \Delta + \Delta \tepro \delta )$ is the co-Poisson structure
naturally associated to the tensor product space (compare to
eqn.(\ref{TP})).
\end{definition}
Notice that these relations are dual to the ones satisfied by
Poisson algebras.

Later in this paper we will define a concept of quantization
for both Poisson and co-Poisson algebras, but first we
will consider so called Poisson Lie groups which
can be interpreted as phase spaces of
classical dynamical systems living on group manifolds
. As we will see these Poisson
Lie groups are closely related to the classical Yang-Baxter
equation and it is them that we will ultimately quantize (
or more accurately their universal enveloping algebras).

\section{Poisson-Lie groups and Lie bi-algebras}
In the approach to quantum groups we consider in this paper
the basic objects are Poisson-Lie groups
\cite{Dr4}. In this section we
study some of their properties and show  how they are
related to Lie bi-algebras and the classical Yang-Baxter equation.
As an example we will consider the groups $SL_N$ which will
also serve as an illustration of the quantization of
Poisson-Lie groups in the next sections.

We start with the definition.
\begin{definition}
A Lie group $G$ is called a Poisson Lie group if the space of
smooth functions on $G$ is a Poisson Hopf algebra.
\end{definition}
Obviously the Hopf algebra structure of
$C^{\infty}(G)$ is the one given in example 1 and
is completely fixed by the structure of the group. It is
therefore the Poisson bracket that has to satisfy a certain
compatibility relation (see eqn. (\ref{comp})) which means
that not every Poisson structure on a Lie group turns it
into a Poisson-Lie group. We will study this compatibility relation
in detail below.

The following lemma gives the general form of a Poisson bracket
on a Lie group.
\begin{lemma}
Let $\{X_{\mu} \}_{\mu = 1}^{dim(G)}$ be a set of right invariant
vectorfields on $G$ (i.e. if $R_g (g')=g'g$, then $X_{\mu} \!\! \mid_g
=(R_g)_{\ast}X_{\mu} \!\! \mid_e $ where $(R_g)_{\ast}$ is the
derivative of $R_g$, and $e$ is the unit element of $G$).
such that $\{X_{\mu}\!\! \mid_g \}$ is a basis in
$T_g G$ for all $g \epsilon G$. Then a Poisson bracket on
$C^{\infty}(G)$ can be written as
\be  \label{Pois}
\{ \phi , \psi \} = \sum_{\mu \nu } \eta^{\mu \nu}(g)
X_{\mu} \!\! \mid_g \! (\phi) \, X_{\nu} \!\! \mid_g \! (\psi)
\ee
where $g \epsilon G$.
\end{lemma}
This follows from the fact that $\{ \phi, .\}: C^{\infty}(G)
\rightarrow C^{\infty}(G)$ is a derivation which means that it
is equal to a vectorfield. Since $\{X_{\mu} \!\! \mid_g \}$ spans
$T_g G$ in every $g \epsilon G$ we can therefore write
$\{ \phi , . \}(g)= \sum_{\mu} \gamma^{\mu}(g) X_{\mu}\!\! \mid_g$.
Applying the same argument to $\{., \psi \}$ the lemma follows.

We can rewrite the  form of this Poisson bracket somewhat:
\begin{eqnarray}
\{\phi , \psi \}(g) & = & \sum_{\mu \nu} \eta^{\mu \nu}(g)
X_{\mu}\!\! \mid_g \! (\phi) \,
X_{\nu} \!\! \mid_g \! (\psi) \nonumber \\
& = & \mbox{} \sum_{\mu \nu} \eta^{\mu \nu}(g) (d \phi \!\! \mid_g
\tepro d \psi \!\! \mid_g )
(X_{\mu} \!\! \mid_g \tepro X_{\nu} \!\! \mid_g) \nonumber
\\
& = & \mbox{} \eta (g)
(d \phi \!\! \mid_g \tepro d \psi \!\! \mid_g) \label{PB}
\end{eqnarray}
where $\eta: G \rightarrow L \tepro L$ is defined by ($L$ is the
Lie algebra of $G$)
\be
g \longmapsto
\eta (g) = \sum_{\mu \nu} \eta^{\mu \nu}(g) X_{\mu}
\tepro X_{\nu}
\ee
and also $d \phi \!\! \mid_g (X_{\mu}) \equiv d \phi \!\! \mid_g
(X_{\mu} \!\! \mid_g )$. Here we have identified the space of
right invariant vectorfields with $T_e G =L$. This can be done
because $X_{\mu} \!\! \mid_e$ determines $X_{\mu}$ in any point
$g \epsilon G$ by right translation (i.e. the fields $X_{\mu}$
are right
invariant which means by definition $X_{\mu} \!\! \mid_g =
(R_g )_{\ast} X_{\mu} \!\! \mid_e$ where  $(R_g)_{\ast}$  is the
derivative  of $R_g$). We can therefore consider  $\{ X_{\mu} \}$
to be a basis of $L$.

Of course the fact that $\{., .\}$ is a Poisson bracket and also that
it is coordinated to the Hopf algebra structure on $C^{\infty}(G)$
gives the map $\eta$ certain properties. This what we will investigate
next.

Let $C^n(G;L)$ be the space of maps
\be
\lambda : G \times \ldots \times G \rightarrow L \tepro L
\ee
We can turn the sequence $\{C^n (G;L) \}_{n=0}^{\infty}$ into
a complex by defining the coboundary operator
\be
\delta_G : C^n(G;L) \rightarrow C^{n+1}(G;L)
\ee
as follows
\begin{eqnarray}
[\delta_G \lambda ](g_1, \ldots , g_{n+1}) & = &
g_1 . \lambda (g_2, \ldots , g_{n+1}) \nonumber \\
&   & \mbox{} +\sum_{i=1}^{n}(-1)^i \lambda (g_1, \ldots , g_i g_{i+1}
, \ldots , g_{n+1}) \nonumber \\
&   & \mbox{} + (-1)^{n+1} \lambda (g_1, \ldots , g_n)
\end{eqnarray}
(where $g_i \epsilon G$ and $\lambda \epsilon C^n(G;L)$). The
action of $G$ on $L \tepro L$ (which we used in the definition)
is defined by
\be
g.(X \tepro Y)=Ad_gX \tepro Ad_gY \equiv Ad_g^{\tepro 2}(X \tepro Y)
\ee
where $Ad$ denotes the adjoint action and $X, Y \epsilon L$. It
is a straightforeward computation to show that $\delta_G^{2}=0$.
The compatibility relation is the subject of the following
theorem.
\begin{theo}
Let $\eta$ be the map associated to the Poisson structure on
$C^{\infty}(G)$ via the relation (\ref{PB}). Then the
compatibility relation (\ref{comp}) between the Poisson bracket
and the Hopf algebra structure on $C^{\infty}(G)$ is equivalent
to the cocycle condition on $\eta$, i.e.
\be
\delta_G \eta =g_1. \eta(g_2)-\eta(g_1 g_2) +\eta (g_1)=0
\ee
\end{theo}
The proof of this theorem is an explicit calculation. If we write
$\Delta(\phi)=\sum_i \phi^{(1)}_i \tepro \phi^{(2)}_i$ then by
definition of the co-product on $C^{\infty}(G)$ (see example 1)
we have
\begin{eqnarray}
\Delta \phi (g_1, g_2) & = & \phi (g_1 g_2) \nonumber \\
& = & \mbox{} \sum_i \phi^{(1)}_i (g_1) \, \phi^{(2)}_i (g_2)
\end{eqnarray}
Also remembering the definition of the Poisson bracket on the
tensor product space $C^{\infty}(G) \times C^{\infty}(G)$
\be
\{ \Delta (\phi), \Delta (\psi) \}=\sum_{ij} \{\phi^{(1)}_i,
\psi^{(1)}_j \} \tepro \phi^{(2)}_i \psi^{(2)}_j
+\phi^{(1)}_i \psi^{(1)}_j \tepro \{ \phi^{(2)}_i, \psi^{(2)}_j \}
\ee
we find
\begin{eqnarray}
\{ \Delta (\phi), \Delta (\psi)\}(g_1, g_2) & = & \sum_{\mu \nu}
\sum_{ij} (\eta^{\mu \nu}(g_1) \, X_{\mu}\!\! \mid_{g_1} \!\!
(\phi_i^{(1)}) \,
X_{\nu} \!\!
\mid_{g_1} \!\! (\psi^{(1)}_j) \, \phi^{(2)}_i (g_2) \psi^{(2)}_j
(g_2) \nonumber \\
&   & \mbox{} +\phi^{(1)}_i (g_1) \psi^{(1)}_j (g_1) (\eta^{\mu \nu}
(g_2)\, X_{\mu} \!\! \mid_{g_2} \!\!
(\phi^{(2)}_i) \, X_{\nu} \!\! \mid_{g_2} \!\!
(\psi^{(2)}_j))
\end{eqnarray}
We also have
\begin{eqnarray}
\sum_i X_{\mu} \!\! \mid_{g_1} \!\!
( \phi^{(1)}_i ) \, \phi^{(2)}_i (g_2) & = &
\frac{d}{dt} \sum_i  \phi^{(1)}_i (e^{t X_{\mu}}g_1) \, \phi^{(2)}_i
(g_2) \mid_{t=0} \nonumber \\
& = & \mbox{} \frac{d}{dt} \phi (e^{tX_{\mu}}g_1 g_2) \mid_{t=0}
=d\phi \!\! \mid_{g_1 g_2}(X_{\mu} )
\end{eqnarray}
In a similar way we can derive
\be
\sum_i \phi^{(1)}_i (g_1) \, X_{\mu} \!\!
\mid_{g_2} \!\! (\phi^{(2)}_i)=
d\phi \!\! \mid_{g_1 g_2}(Ad_{g_1}X_{\mu} )
\ee
With these results we arrive at
\be
\{ \Delta (\phi), \Delta (\psi) \} (g_1, g_2) =
d\phi \!\! \mid_{g_1 g_2} \tepro
d\psi \!\! \mid_{g_1 g_2} (\eta (g_1) + g_1 .
\eta (g_2))
\ee
By definition we have however
\be
\Delta \{ \phi , \psi \} (g_1 , g_2 )= (d \phi \!\!
 \mid_{g_1 g_2} \tepro
d\psi \!\! \mid_{g_1 g_2}). \eta (g_1 g_2)
\ee
Equating these two relations we get the desired result.

So for the Poisson bracket to be coordinated to the Hopf
structure the map $\eta$ must be a 2-cocycle. This gives us
one of the properties of $\eta$. The other properties, i.e the
ones associated to the anti-symmetry and the Jacobi-identity
are now easily deduced.

Associated to the map $\eta$ we define
\be
\phi_{\eta}: L \rightarrow L \tepro L
\ee
by
\be  \label{colie}
\phi_{\eta}(X)=\frac{d}{dt} \eta (e^{tX})\mid_{t=0}
\ee
One might call $\phi_{\eta}$ the infinitesimal version of $\eta$
in the unit element of $G$. The map $\phi_{\eta}$ inherits from
$\eta$ certain properties which are listed in the following
theorem.
\begin{theo}
Let $\eta$ be the map associated to the Poisson structure on
the Poisson Lie group $G$
and let $\phi_{\eta}$
be defined by eqn.(\ref{colie}), then
\begin{enumerate}
\item $\phi_{\eta}$ is co-antisymmetric.
\item $\phi_{\eta}$ satisfies the co-Jacobi identity.
\item $\phi_{\eta}([X, Y])=X.\phi_{\eta}(Y)-Y.\phi_{\eta}(X)$
\end{enumerate}
where $L$ acts on $L \tepro L$ via $X.(Y \tepro Z)=
[X, Y] \tepro Z + Y \tepro [X, Z]$ which is the infinitesimal
version of the action of $G$ on $L \tepro L$.
\end{theo}
Proof: Anti-symmetry of the Poisson bracket gives
\begin{eqnarray}
\eta (d\phi \tepro d \psi) & = & \sum \eta^{\mu \nu}(g)
X_{\mu}(\phi) X_{\nu}(\psi) \nonumber \\
& = & \mbox{} -\sum \eta^{\mu \nu}(g) X_{\mu} (\psi) X_{\nu}(\phi)
\nonumber \\
& = & \mbox{} -(\tau \circ \eta) (d \phi \tepro d\psi)
\end{eqnarray}
so indeed we find $\eta = -\tau \circ \eta$. Using the definition
of $\phi_{\eta}$ we see that $\tau \circ \phi_{\eta}=-\phi_{\eta}$.
The proof that $\phi_{\eta}$ satisfies the co-Jacobi identity
is similar. The third property is slightly trickier. First note
that from the co-cycle condition follows $\eta (e)=0$
(insert $g_1=e$ into the cocycle condition). Therefore
\begin{eqnarray}
0 & = & \partial_t \eta (e^{tX} e^{-tX})\mid_{t=0} \nonumber \\
  & = & \mbox{} \partial_t \eta (e^{tX}) \mid_{t=0}+
\partial_t ((Ad^{\tepro 2}_{e^{tX}}) \eta (e^{-tX})) \mid_{t=0}
\nonumber \\
 & = & \phi_{\eta}(X) + \phi_{\eta}(-X) \label{nul}
\end{eqnarray}
Then we have
\begin{eqnarray}
\phi_{\eta} ([X, Y]) & = & \frac{d}{ds}\frac{d}{dt} \eta (
e^{sX}e^{tY}e^{-sX} \mid_{t=0} \nonumber \\
& = & \frac{d}{ds} \frac{d}{dt} (\eta (e^{tX})+(Ad^{\tepro 2}_{
e^{sX}}) \eta (e^{tY}) +Ad^{\tepro 2}_{e^{tX}} Ad^{\tepro 2}_{e^{tY}}
\eta (e^{-sX}) \mid_{s, t=0} \nonumber \\
& = & \frac{d}{ds} (Ad^{\tepro 2}_{e^{sX}}) \mid_{s=0}
\phi_{\eta} (Y)+ \frac{d}{dt} (Ad^{\tepro 2}_{e^{tY}}) \mid_{t=0}
\phi_{\eta}(-X) \nonumber \\
& = & ad_X^{\tepro 2} \phi (Y)+ ad_Y^{\tepro 2} \phi(-X) \nonumber \\
& = & X. \phi_{\eta}(Y) - Y. \phi_{\eta}(X)
\end{eqnarray}
where we used eqn.(\ref{nul}) in the last step. This proves the
theorem.

In general we call a Lie algebra $L$ together with a map
$\phi :L \rightarrow L \tepro L$ such that the properties
1, 2 and 3  stated in the theorem are satisfied a Lie bi-algebra.
 From the foregoing follows that the Lie algebra of
a Lie Poisson group is a Lie bi-algebra.

A trivial way to satisfy the co-cycle condition $\delta_G \eta =0$
is of course to choose $\eta$ to be a coboundary
\be
\eta= \delta_G r
\ee
for some $r \epsilon L \tepro L$. Let us calculate the Lie
co-bracket $\phi_{\eta}$ associated to such an $\eta$. From the
definition of $\delta_G$ we find
\be
[\delta_G r](g)=r-g.r
\ee
so we get
\begin{eqnarray}
\phi (X) & = & \frac{d}{dt} \eta (e^{tX}) \mid_{t=0}=\frac{d}{dt}
(r-e^{tX}.r) \mid_{t=0} \nonumber \\
& = & \mbox{} -X.r
\end{eqnarray}
Write for the moment $r=r^{\mu \nu}X_{\mu} \tepro X_{\nu}$, then
\begin{eqnarray}
\phi (X) & = & -r^{\mu \nu}([X, X_{\mu}] \tepro X_{\nu} +
X_{\mu} \tepro [X, X_{\nu}]) \nonumber \\
& = & \mbox{} [r, 1 \tepro X + X \tepro 1]
\end{eqnarray}
Such a choice for $\eta$ (and $\phi$) trivially satisfies the
cocycle condition, which was related to the fact that the
Poisson structure and the Hopf structure are related.
The co-antisymmetry and co-Jacobi identities for $\phi$ have not
yet been considered. Obviously these will restrict the possible
choices for $r$. We give the conditions on $r$ in a theorem.
\begin{theo}
Let $L$ be a Lie algebra and $r$ an element of $L \tepro L$.
Choose an arbitrary basis $\{ X_{\mu} \}$ in $L$
and write $r=r^{\mu \nu}X_{\mu} \tepro X_{\nu}$. Also define
\begin{eqnarray}
r_{+} & = & \frac{1}{2}(r^{\mu \nu} + r^{\nu \mu}) X_{\mu}
\tepro X_{\nu} \\
r_{-} & = & \frac{1}{2}(r^{\mu \nu} - r^{\nu \mu})
X_{\mu} \tepro X_{\nu} \\
r_{12} & = & r^{\mu \nu}X_{\mu} \tepro X_{\nu} \tepro 1  \\
r_{13} & = & r^{\mu \nu}X_{\mu} \tepro 1 \tepro X_{\nu}  \\
r_{23} & = & r^{\mu \nu}1 \tepro X_{\mu} \tepro X_{\nu}
\end{eqnarray}
Then the map $\phi : L \rightarrow L \tepro L$ defined
by
\be
\phi (x) = [ r, X \tepro 1+1 \tepro X ]
\ee
turns $L$ into a Lie bi-algebra if and only if
\begin{itemize}
\begin{enumerate}
\item $r_{+}$ is ad-invariant, i.e. $(Ad_g \tepro Ad_g)r_+=r_+$
\item $B=[r_{12}, r_{13}]+[r_{13}, r_{23}]+[r_{12}, r_{23}]$ is
ad-invariant, i.e. $(Ad_g \tepro Ad_g \tepro Ad_g) B=0$
\end{enumerate}
\end{itemize}
for all $g \epsilon G$. $B$ is called the schouten bracket of
$r$ with itself.
\end{theo}
The proof of this theorem is a lengthy but straightforward
calculation.  One has to write
out the co-antisymmetry and co-Jacobi identities for
the map $\phi$ given above.

The equation $Ad_g^{\tepro 3} B=0$ is called the
modified (classical) Yang-Baxter equation while the equation
$B=0$ is simply called the classical Yang-Baxter equation \cite{sem}.
As we will see	later the classical Yang-Baxter equation
is the classical limit of the even more important quantum
Yang-Baxter equation (also known in statistical mechanics
as the star-triangle equation).

Let us consider an example.
\begin{example}
Let $L=sl_N$ and let $r$ be given by
\be
r=C-\sum_{i<j} e_{ij} \wedge e_{ji}
\ee
where
\be
C=\sum_{i \neq j} e_{ij} \tepro e_{ji} +\sum_{\mu \nu =1}^{N-1}
K^{\mu \nu}H_{\mu} \tepro H_{\nu}
\ee

and $(e_{ij})_{kl}=\delta_{ik}\delta_{jl}$. $H_{\mu}$ are
the standard generators of the
Cartan subalgebra. $K^{\mu \nu}$ is the inverse of the Cartan matrix.

In the special case of $sl_2$ this reduces to
\be
r=\frac{1}{2}H \tepro H +2 E \tepro F
\ee
such that $\phi$ becomes
\begin{eqnarray}
\phi (H) & = & 0 \\
\phi (E) & = & \frac{1}{2} E\wedge H \\
\phi (F) & = & \frac{1}{2} F \wedge H
\end{eqnarray}
This map will play a crucial role in the quantization later
on.
\end{example}
We can write the Poisson bracket (\ref{Pois}) explicitly in terms
of $r$. Since $\eta (g)=r- Ad_g r$ we get
\begin{equation}
\eta^{\mu \nu}(g) X_{\mu} \!\! \mid_g \tepro X_{\nu} \!\! \mid_g
=r^{\mu \nu} (X_{\mu} \!\! \mid_g \tepro X_{\nu} \!\! \mid_g -
Ad_g X_{\mu} \!\! \mid_g \tepro Ad_g X_{\nu} \mid_g )
\end{equation}
Now using $X_{\mu} \!\! \mid_g =(R_g)_{\ast} X_{\mu} \!\! \mid_e$
and $Ad_g=(L_g)_{\ast} (R_g^{-1})_{\ast}$ we find $Ad_g X_{\mu}
\!\! \mid_g = (L_g)_{\ast}X_{\mu} \!\! \mid_e $. Denoting
$(R_g)_{\ast} X_{\mu} \!\! \mid_e (\phi )$ by $\partial_{\mu} \phi$
and $(L_g)_{\ast} X_{\mu} \!\! \mid_e$ by $\partial_{\mu}' \phi$
we get
\be
\{f, g\}=r^{\mu \nu}(\partial_{\mu} f\partial_{\nu}g -
\partial^{\prime}_{\mu}f \partial^{\prime}_{\nu}g)
\ee

As we said earlier a Poisson structure on $C^{\infty}(G)$
induces a co-Poisson structure on ${\cal U}(L)$ because of the
duality between these two spaces. We conclude that the
universal enveloping algebra of a Poisson Lie group $G$ is a
co-Poisson Hopf algebra. Denote the co-Poisson structure on
${\cal U}(L)$ by $\delta$. We then have the following theorem
which gives the relation between $\delta$ and $\phi_{\eta}$.
\begin{theo}
The restriction of $\delta$ to $L \subset {\cal U}(L)$
is equal to $\phi_{\eta}$.
\end{theo}

Proof: Denote the Poisson structure (\ref{Pois}) by $\gamma$,
i.e. $\{ \phi , \psi \} = \gamma (\phi \tepro \psi )$. The map
$\delta$ is defined by the relation
\be
\langle \gamma (\phi \tepro \psi ), a \rangle =
\langle \phi \tepro \psi , \delta (a) \rangle
\ee
where $\phi , \psi \epsilon C^{\infty}(G)$, $a \epsilon {\cal U}
(L)$ and $\langle ., . \rangle $ denotes the duality  between
$C^{\infty}(G)$ and ${\cal U}(L)$. Denoting $X_{\mu} \phi$
by $\partial_{\mu} \phi$ the left hand side of this equation
is equal to
\begin{eqnarray}
\sum_{\mu \nu} \langle \eta^{\mu \nu} \partial_{\mu} \phi
\partial_{\nu}\psi , X \rangle & = &
\rho (X) (\eta^{\mu \nu} \partial_{\mu} \phi  \partial_{\nu}
\psi )(e) \nonumber \\
& = & \mbox{} \frac{d}{dt} \eta^{\mu \nu}(e^{tX})
\, \partial_{\mu} \phi (e^{tX}) \, \partial_{\nu} \psi (e^{tX})
\mid_{t=0} \nonumber \\
& = & \frac{d}{dt} \eta^{\mu \nu}(e^{tX}) \mid_{t=0}
\partial_{\mu} \phi (e) \, \partial_{\nu} \psi (e) \nonumber \\
& = & \phi_{\eta}^{\mu \nu}(X) X_{\mu} \!\! \mid_e \!\!(\phi) \,
X_{\nu} \!\! \mid_e \!\! (\psi )
\end{eqnarray}
for $X \epsilon L \subset {\cal U}(L)$. Since
\be
\langle \phi, X \rangle = \frac{d}{dt} \phi (e^{tX}) \mid_{t=0}
=X \!\! \mid_e \!\! (\phi) = d \phi \!\! \mid_e \!\! (X)
\ee
we find $\langle \phi \tepro \psi , \delta (X) \rangle =
(d\phi \!\! \mid_e \tepro d\psi \!\! \mid_e )\delta (X) \rangle$
for the right hand side. The theorem follows.

Quantizing Poisson Lie groups can  be
performed at different levels. One could quantize the Poisson
Hopf algebra $C^{\infty}(G)$, or equivalently the
co-Poisson Hopf algebra ${\cal U}(L)$. In the next section
we will define quantization for both, however we will only
persue the quantization of ${\cal U}(L)$.

\section{Deformation quantization of (co)-Poisson structures}
In section 2 we defined (co)-Poisson algebras and in section 3
we saw how they are related to Poisson Lie groups. It is the
purpose of this section to define a suitable form of quantization,
called deformation quatization \cite{BFFLS},
for these objects. Using  this
definition we will quantize the universal enveloping algebra
of $sl_2$, which is a (co)-Poisson Hopf algebra as we saw in the
previous section.

\begin{definition}
Let $(A_0, m_0, \eta_0, \{., .\})$ be a Poisson algebra over the complex
numbers ${{\bf C}}$. A quantization of $A_0$ is a non-
commutative algebra $(A, m, \eta )$ over the ring ${{\bf C}}[[\hbar]]$
, where $\hbar$ is a formal parameter (interpreted as
Planck's constant), such that
\begin{enumerate}
\item $A/\hbar A \cong A_0$
\item $m_0 \circ (\pi \tepro \pi )= \pi \circ m$
\item $\pi \circ \eta =\eta_0$
\item $\{ \pi (a), \pi (b) \}=\pi (\frac{
[a, b]}{\hbar})$ for all $a, b \epsilon A$.
\end{enumerate}
where $\pi$ denotes the
canonical quotient map
\be
\pi:A \rightarrow A/\hbar A \cong A_0
\ee
and $[., .]$ denotes the ordinary commutator (with respect to
the multiplication $m$).

\end{definition}
Let us consider  this definition more closely. First of all
, because of property 1, the space $A$ can be seen as the
set of polynomials in $\hbar$ with coefficients in $A_0$.
Factoring out $\hbar A$ is then equivalent to putting
$\hbar =0$ which corresponds to the classical limit.
However the multiplication on $A$ is not simply the
multiplication which we would have obtained had we
extended the multiplication on $A_0$ to $A$ (i.e.
if $a=\sum_{i=0}^{\infty}a_i \hbar^i$ and
$b=\sum_{j=0}^{\infty} b_j \hbar^j$ then
naively extending the multiplication of $A_0$ to $A$ would
have given $a.b=\sum_{ij}a_i b_j \hbar^{i+j}$. This
multiplication however is commutative since the multiplication
on $A_0$ in commutative by definition of a Poisson algebra).
On $A$ there is a new, non-commutative, multiplication $m$ which
we denote by $\star$ (i.e. $m(a \tepro b) \equiv a \star b$).
Quite generally one can say that this multiplication is
of the form
\be  \label{mult}
a \star b = \sum_{n=0}^{\infty} f_n (a, b)\hbar^n
\ee
where $f_n:A \times A \rightarrow A$. Property 2 of a quantization
is equivalent to saying that in the classical limit ($\hbar
\rightarrow 0$) $m$ must reduce to $m_0$ (i.e. $\star$ must reduce
to $.$). This fixes $f_0(a, b)=a.b$. Also $m$ must be
associative
\be
(a \star b) \star c=a \star (b \star c)
\ee
which leads to the identity
\be
\sum_{n=0}^{\infty} f_n(f_{l-n}(a, b), c)=
\sum_{n=0}^{\infty} f_n (a, f_{l-n}(b, c))
\ee
(for all $l$). Furthermore since
 $1 \star a = a \star 1 =a$ we have
\be
f_n(a, 1)=f_n(1, a)=a \delta_{n, 0}
\ee

Another point that needs to be cleared up is the
$1/\hbar$ factor in the right hand side of property 2,
since one might say that this could cause terms with
negative powers of $\hbar$ which are not in the algebra $A$.
That everything is alright is contained in the
following lemma.
\begin{lemma}
The commutator $[a, b]$ is at least of order $\hbar$ for
all elements $a, b$ of $A$.
\end{lemma}
The proof is easy and goes as follows: We know that
since $A_0$ is commutative, $0=[\overline{a}, \overline{b}]$.
By property 1 the equivalence class $\overline{a}$ is given by
$a+\hbar A$. Therefore $0=[a+\hbar A, b+\hbar A]=[a, b]+\hbar A$.
Again by property 1 this is equal to $\overline{[a, b]}$
which means that $[a, b]$ is in the kernel of $\pi$ which
is equal to $\hbar A$. This proves the lemma.

As far as we know there is no general theorem stating that every
Poisson algebra can be quantized in the way descibed above or
that a quantization is unique if it exists.

Having defined deformation quantization for Poisson algebras we
can dualize this definition in order to get a definition of
quantization for co-Poisson bi-algebras. Obviously the dual
analogue of a commutator $[., .]$ ($=m-m \circ \tau $) is the
map $\Delta - \tau \circ \Delta$. Motivated by this we come to
the following definition.

\begin{definition}
Let $(A_0, m_0, \Delta_0, \eta_0, \varepsilon_0;\delta)$
be a co-Poisson bi-algebra where $\delta$ denotes the co-Poisson
structure.
A quantization of $A_0$ is a non co-commutative
bi-algebra $(A, m, \Delta, \eta, \varepsilon )$
over the ring ${{\bf C}}[[\hbar]]$ such that
\begin{enumerate}
\item $A/\hbar A \cong A_0$
\item $(\pi \tepro \pi ) \circ \Delta = \Delta_0 \circ \pi$
\item $m_0 \circ (\pi \tepro \pi )= \pi \circ m$
\item $\pi \circ \eta = \eta_0$
\item $\varepsilon \circ \pi = \varepsilon_0$
\item $\delta(\pi(a))=\pi(\frac{1}{\hbar}(\Delta(a)
-\tau \circ \Delta (a))$ for all $a$ in $A$
\end{enumerate}
where $\pi$ again denotes the canonical quotient map $\pi:A
\rightarrow A_0$.
\end{definition}

W.r.t. this definition similar remarks can be made as before
(we will not repeat them). We do want to bring another point to
the attention of the reader. Even though the quantization of a
co-Poisson structure involves  primarily the comultiplication
(see point 6 in the definition) the other structures in the
bi-algebra may also be deformed (i.e. the maps $\varepsilon, m, \eta$
will not simply be the maps $\varepsilon_0, m_0, \eta_0$ extended to
$A$) because in a bi-algebra the different maps are coordinated.
The axiom relating the multiplication to the co-multiplication
reads
\be
\Delta_0 (a_0.b_0) =\Delta_0 (a_0) .\Delta (b_0)
\ee
and
\be  \label{comult}
\Delta (a \star b ) =\Delta (a) \star \Delta (b)
\ee
in $A_0$ and $A$ respectively. From eqn.(\ref{comult}) one
immediately sees that given $\Delta$ the multiplication $m_0$ may
have to be altered in order to satisfy this relation. The same can
be said about the other
structure maps because of their relations with the (co)-multiplication.

In the next section we will consider in detail the quantization
of the universal enveloping algebra of $sl_2$ which is a
co-Poisson Hopf algebra. Indeed we will find that not only the
co-multiplication is deformed but all the other Hopf algebra
structures as well.

\section{The quantization of ${\cal U}(sl_2)$}
In this section we will undertake the quantization of
the universal enveloping algebra of $sl_2$ equipped
with the co-Poisson structure discribed in section 3.
What we will obtain is a non-commutative and non-cocommutative
Hopf algebra called the quantized universal enveloping algebra
denoted by ${\cal U}_q (sl_2)$. This algebra was first introduced
by Drinfeld and in a slightly different form (and starting from
a different principle) by Jimbo.

As we saw in section 3 the co-Poisson structure of
${\cal U}(sl_2)$ is given by the extension to the
entire universal enveloping algebra of the map
\begin{eqnarray}
\delta (H) & = & 0  \\
\delta (E) & = & \frac{1}{2}E \wedge H	\\
\delta (F) & = & \frac{1}{2}F \wedge H
\end{eqnarray}
(the Lie algebra $sl_2$ itself together with this this map
was called a Lie bi-algebra). As a space the quantization
of the universal enveloping algebra is known, as we saw in
the previous paragraph. It is simply the set of (formal)
polynomials
in $\hbar$ with coefficients in ${\cal U}(sl_2)$. What we
have to do first is find the coproduct $\Delta$ on this new space.
This co-product is determined by the following requirements:
\begin{enumerate}
\item $\Delta$ must be co-associative (or else the quantized
algebra will not be a Hopf algebra).
\item $\delta (\pi(a))=\pi(\frac{1}{\hbar}
(\Delta (a)- \tau \circ \Delta (a)))$
\item In the classical limit ($\hbar \rightarrow 0$) the
coproduct $\Delta$ must reduce to the ordinary coproduct on
${\cal U}(sl_2)$.

\end{enumerate}

The comultiplication $\Delta$ has the general form
\be
\Delta = \sum_{n=0}^{\infty} \frac{\hbar^n}{n!}\Delta_{(n)}
\ee
The third requirement fixes
$\Delta_{(0)}$:
\begin{eqnarray}
\Delta_{(0)}(H) & = & H \tepro 1 + 1 \tepro H \\
\Delta_{(0)}(E) & = & E \tepro 1 + 1 \tepro E \\
\Delta_{(0)}(F) & = & F \tepro 1 + 1 \tepro F
\end{eqnarray}
The second requirement reads as follows:
\begin{eqnarray}
\delta (H) & = & 0 =\Delta_{(1)}(H)-\tau \circ \Delta_{(1)}(H) \\
\delta (E) & = & \frac{1}{2}E \wedge H=\Delta_{(1)}(E)-
\tau \circ \Delta_{(1)} (E) \\
\delta (F) & = & \frac{1}{2}F \wedge H=\Delta_{(1)}(F)-
\tau \circ \Delta_{(1)} (F)
\end{eqnarray}
An obvious solution of this system of equations is given by:
\begin{eqnarray}
\Delta_{(1)}(H) & = & 0 \\
\Delta_{(1)}(E) & = & \frac{1}{4}E \wedge H \\
\Delta_{(1)}(F) & = & \frac{1}{4}F \wedge H
\end{eqnarray}
So using the classical limit and the quantization condition
for the Poisson structure we have been able to find
the two lowest order elements of $\Delta$. In order to
find the higher order terms we can use requirement 1
which states that $\Delta$ must be co-associative. Inserting
the expansion of $\Delta$ into the equation which expresses
the coassociativity and collecting terms of order $\hbar^n$
we come to the following recursive relation:
\be \label{rec}
\sum_{k=0}^{n}\left(\begin{array}{c} n\\ k \end{array} \right)
(\Delta_{(k)}\tepro 1-1 \tepro
\Delta_{(k)}) \Delta_{(n-k)}=0
\ee
If we know all the $\Delta_{(k)}$ for $k <n$ we get an
equation for $\Delta_{(n)}$. In this way we can solve
the above equation recursively (remember we already know
$\Delta_{(0)}$ and $\Delta_{(1)}$). It is
now easy to show by induction that for arbitrary $n$
\begin{eqnarray}
\Delta_{(n)}(H) & = & 0 \\
\Delta_{(n)}(E) & = & \frac{1}{4^n}(E \tepro H^n +(-1)^n
H^n \tepro E) \\
\Delta_{(n)}(F) & = & \frac{1}{4^n}(F \tepro H^n +(-1)^n
H^n \tepro F)
\end{eqnarray}
solve the recursion relation for all $n$. Here by
$H^n$ is meant $H \star H \star ...\star H$ (n-times) where
$\star$ is the deformed multiplication  on $A$.
First one has to show that $\Delta_{(n)}(H)=0$ solves eqn.(\ref{rec})
(this is easy).
In the induction step for $E$ and $F$ we need to know what
$\Delta_{(N)}(H \star H)$ is. This can be derived from the fact
that the quantized algebra must still be a Hopf algebra which
implies the identity $\Delta (H \star H)=\Delta (H) \star
\Delta (H)$. The l.h.s. of this equation is equal to
$\sum_n \frac{\hbar^n}{n!} \Delta_{(n)}(H \star H)$ while the
r.h.s. is equal to $\sum_{nm}\frac{h^{n+m}}{n!m!}
\Delta_{(n)}(H) \star \Delta_{(m)}(H)=\Delta_{(0)}(H) \star
\Delta_{(0)}(H)$. Therefore $\Delta_{(n)}(H \star H)=0$
for $n>0$ and $H^2 \tepro 1 + 2H \tepro H + 1 \tepro H^2$ for
$n=0$.

Using these results and the expansion of $\Delta$ we
find the co-product of the quantized universal enveloping
algebra to be
\begin{eqnarray}
\Delta (H) & = & H \tepro 1 + 1 \tepro H \\
\Delta (E) & = & E \tepro q^H+q^{-H} \tepro E \\
\Delta (F) & = & F \tepro q^H +q^{-H} \tepro F
\end{eqnarray}
where $q=e^{\hbar /4}$.

Let us pause for a moment to reflect the result.
First of all note that in the limit $\hbar \rightarrow 0$
we indeed recover the old co-multiplication of the universal
enveloping algebra. Also note that the co-multiplication
of the Cartan element is not deformed. On the whole
 we have found a non co-commutative co-product. As
we will see however the non-commutativity of this
co-product is under control, i.e. there does exist a relation
between $\Delta$ and $\tau \circ \Delta$. Therefore the
tensorproduct representations $V\tepro W$ and $W \tepro V$
are no longer equal but still equivalent. We will come
to this later.

Using the definition of quantization we have found the
co-multiplication of the quantized universal enveloping
algebra of $sl_2$. What about the other structures that
make the universal enveloping algebra into a Hopf algebra?
As we discussed in the previous section the
structure maps of a Hopf algebra
are coordinated, and changing one of these
maps, even by a small amount,
 may violate the Hopf algebra axioms.
Therefore we have to check if the co-multiplication
we found for the quantized universal enveloping algebra
(QUEA) is still compatible with the other structure maps
on the UEA. If not we have to deform the other structure
maps in such a way that the totally deformed Hopf algebra
still is a Hopf algebra and reduces to the ordinary
UEA in the classical limit.

First consider the multiplication map. One of the axioms
of a Hopf algebra is that for all $a, b$ we must have
\be
\Delta (a \star b)=\Delta (a) \star \Delta (b)
\ee
In particular this means that the following equalities must
hold
\begin{eqnarray}
\Delta ([H, E])&=&[\Delta (H), \Delta (E)] \\
\Delta ([H, F])&=&[\Delta (H), \Delta (F)] \\
\Delta ([E, F])&=&[\Delta (E), \Delta (F)]
\end{eqnarray}
It is easily verified that if we take
$[H, E]=2E$ and $[H, F]=-2F$ then the first two relations
are satisfied. Therefore these commutation relations,
which are the ordinary $sl_2$ relations,
are still consistent with the Hopf algebra axioms even
after quantization. The situation is different with the
third relation however. Working out the right hand side
of this relation we get
\be
\Delta ([E, F])=[E, F]\tepro q^{2H}+q^{-2H}\tepro [E, F]
\label{comm}
\ee
It is obvious that $[E, F]=H$ is not a solution of this
equation because $\Delta (H)=H \tepro 1 + 1\tepro H$. In
finding a solution we also have to remember that in the
classical limit we do have to find the relation $[E, F]=H$
back. A commutation relation that satisfies all the
requirements is
\be
[E, F]=\frac{q^{2H}-q^{-2H}}{q-q^{-1}}
\ee
It is an easy exercise to check relation (\ref{comm})
and the classical limit for
this commutation relation.

In the same manner we can proceed with the other structure
maps by confronting them with the coordinating axioms. The
resulting Hopf algebra looks like this
\begin{eqnarray}
[H, E] & = &2E \\{}
[H, F] & = &-2F \\{}
[E, F] & = &[H]_q \\
\Delta (H) & = &H \tepro 1+1\tepro H \\
\Delta (E) & = &E \tepro q^{H}+q^{-H}\tepro E \\
\Delta (F) & = &F \tepro q^{H}+q^{-H}\tepro F \\
\varepsilon (E) & = &\varepsilon (F)=\varepsilon (H)=0 \\
\varepsilon (1) & = &1 \\
S(E) & = &-qE \\
S(F) & = &-q^{-1}F \\
S(H) & = &-H
\end{eqnarray}
where we defined
\be
[x]_q=\frac{q^x -q^{-x}}{q-q^{-1}}
\ee
This Hopf algebra is called the quantum universal enveloping
algebra of $sl_2$ and is denoted by ${\cal U}_q (sl_2)$.

Before moving on a remark is in order. In the above derivation
we have carefully evaded all issues of uniqueness.
To our knowledge there is
no rigorous proof of the uniqueness of the quantization of
a given co-Poisson algebra. What the above construction does
show however is that in the case of the UEA of $sl_2$ there
does \underline{exist} a quantization.

The same procedure can be repeated for the other simple algebras.
The resulting QUEAs have the same deformation structure as
in the $sl_2$ case.

In the next section we will consider an application of
these QUEAs. This will lead us to the so called
quasitriangular Hopf algebras which play an important
role in relation with the quantum Yang-Baxter equation.

\section{Quantum groups and the Yang-Baxter equation}
In this section we we will consider the construction of
solutions of the quantum Yang-Baxter equation using the
quantum group ${\cal U}_q (sl_2)$. The construction we use
is called the quantum double construction which has been applied
successfully to derive the quantum $R$-matrices for many
quantum algebras \cite{Dr1,Ros1,Bur2}. Before we come to
the quantum double construction we consider its classical
analogue.

\subsection{The classical double }

In this section we consider the classical double construction
which allows one to construct a very simple solution of the
classical Yang-Baxter equation on a Lie algebra that is
constructed out of a bi-algebra an its dual
. We do this because the classical
double construction is very easy and will give the
reader a good idea of what the quantum double is all about.

Again consider a Lie bi-algebra $(L, [., .], \phi)$. The
axioms of the map $\phi$ are such that the bracket
$[., .]^{\ast}:L^{\ast} \times L^{\ast} \rightarrow L^{\ast}$
defined by
\be
[f, g]^{\ast}=(f \tepro g) \circ \phi
\ee
(where $f, g \epsilon L^{\ast}$) turns the dual $L^{\ast}$
of $L$ into a Lie algebra. Consider now the vectorspace
$D=L \oplus L^{\ast}$ on which we define the scalar product
\be
\langle (X, f), (Y, g) \rangle=f(Y)+g(X)
\ee
(where $f, g \epsilon L^{\ast}$ and $X, Y \epsilon L$). This
space has some very nice properties one of which is
contained in the following theorem.
\begin{theo}
There exists a unique Lie algebra structure on $D$  such that
\begin{itemize}
\item $L$ and $L^{\ast}$ are Lie subalgebras of $D$.
\item $\langle [A, B], C \rangle = \langle A, [B, C] \rangle$
for all $A, B, C \epsilon D$.
\end{itemize}
\end{theo}
The only non-trivial part of the proof is defining the bracket
$[X, f]$ for $X \epsilon L$ and $f \epsilon L^{\ast}$ such
that the second
property  holds. Imposing this property we get the
following equations:
\be
\langle [X, f], Y \rangle = -\langle f, [X, Y] \rangle \equiv
-(ad^{\ast}_{X}f)(Y)
\ee
and
\be
\langle [X, f], g \rangle = -\langle X, [f, g] \rangle
=(f \tepro g) \phi (X)
\ee
which hold for all $Y \epsilon L$ and $g \epsilon L^{\ast}$.
 From this it follows immediately that
\be
[X, f]=-ad^{\ast}_{X}f +(f \tepro 1)\circ (X)
\ee
The only thing left is to check the axioms of a Lie algebra
for this bracket. This however	is an easy exercise.

If we choose a particular basis $\{X_j\}$ in $L$ and equip
$L^{\ast}$ with the dual basis $\{f^i\}$ (i.e.
$f^i (X_j) = \delta ^{i}_{j}$) then we can easily
verify that the bracket on $D$ discribed in the theorem
above can be written as follows:
\begin{eqnarray}
[X_i, X_j] & = & C^k_{ij}X_k \\{}
[f^i, f^j] & = & \Gamma^{ij}_k f^k \\{}
[f^i, X_j] & = & C^i_{jk} f^k -\Gamma^{ik}_j X_j
\end{eqnarray}
The space $D$ together with this Lie algebra structure and
the scalar product is called the  Double Lie algebra
associated to the Lie bi-algebra $(L, [., .], \phi)$.

The nice thing of a double Lie algebra is that it is
very easy to construct solutions of the classical Yang-Baxter
equation on it. The precise statement of this fact is contained
in the following theorem.
\begin{theo}
The element $r=X_i \tepro f^i \epsilon D \tepro D $, called
the canonical element, is a solution of the classical
Yang-Baxter equation, i.e.
\be
B=[r_{12}, r_{13}]+[r_{12}, r_{23}]+[r_{13}, r_{23}]=0
\ee
\end{theo}
The proof is very easy using the commutation relations
in the double algebra:
\be
B=C^{k}_{ij} X_k \tepro f^i \tepro f^j +C^{i}_{jk} X_i
\tepro f^k \tepro f^j-\Gamma^{ik}_{j} X_i \tepro X_k \tepro
f^j +\Gamma^{ij}_{k} X_i \tepro X_j \tepro f^k =0
\ee
which proves the theorem.

Given a Lie bi-algebra it is therefore straightforward  to
construct a solution of the classical Yang-Baxter equation
on its double algebra. The construction we will use in the
quantum case is a direct quantum analogue of the classical
double construction outlined above.

\subsection{Quasi-triangular Hopf algebras}
In this section we will be taking a look at the so called
(quasi)-triangular Hopf algebras (QTHA)
\cite{Dr1,Majid}. As we will see
QTHAs are closely related to the
quantum Yang-Baxter equation. The method we consider in the
next section for the construction of solutions of the QYBE
makes explicit use of
them. The definition of a QTHA is
\begin{definition}
Let $(A, m, \Delta, \eta, \varepsilon, S)$ be a Hopf algebra
and $R$ an invertible element of $A \tepro A$, then the
pair $(A, R)$ is called a QTHA if
\begin{enumerate}
\item $\Delta ^{\prime}(a)=R\Delta (a) R^{-1}$
\item $(\Delta \tepro 1)R=R_{13}R_{23}$
\item $(1 \tepro \Delta)R=R_{13}R_{12}$
\end{enumerate}
where $\Delta^{\prime}=\tau \circ \Delta$ is the so called
opposite comultiplication.
\end{definition}
If we consider some matrix representation $T=(t_{ij})$ of $A$ then
relation 1 becomes
\be
{\cal R} (\rho \tepro \rho )\Delta '(a) = (\rho \tepro \rho )\Delta (a)
{\cal R}
\ee
where ${\cal R }=(\rho \tepro \rho ) R$. $(\rho \tepro \rho )\Delta
= (1 \tepro T)(T \tepro 1)$ while $(\rho \tepro  \rho)\Delta '$ can
be written as $(T \tepro 1)(1 \tepro T )$. Since $\Delta '$ is
non-cocommutative the elements $t_{ij}$ of $T$ do not commute
in general which means that $(T \tepro 1)(1 \tepro T) \neq
(1 \tepro T)(T \tepro 1)$. If we write $\tilde{T} = T \tepro 1$
and $\tilde{ \tilde{T}}= 1 \tepro T$ then property 1 implies
\be
{\cal R} \tilde{T} \tilde{\tilde{T}}= \tilde{\tilde{T}} \tilde{T}
{\cal R}
\ee
which gives certain permutation relations for the matrix elements of
$T$. This equation  plays an important role in the theory of
solvable lattice models as well as in the quantum inverse scattering
method for integrable quantum field theories.

In relation to the approach we took with respect to Hopf algebras
in the first section we can say the following about this
definition. In contrast to ordinary Hopf algebras these
QTHA have the important property that if $V_1$ and $V_2$
are representations then the tensor product representations
$V_1 \tepro V_2$ and $V_2 \tepro V_1$  are isomorphic, the
isomorphism being provided by the element $R$. This follows
from the fact that the tensor product representation
$V_1 \tepro V_2$ is related to the coproduct $\Delta$ while
the tensorproduct representation $V_2 \tepro V_1$ is related
to the opposite comultiplication $\Delta^{\prime}$. In
an arbitrary Hopf algebra $\Delta$ and $\Delta^{\prime}$ are
not related, however in a QTHA they are related by the
invertible element $R$. This then implies the equivalence
stated above.

We come now to the relevance of QTHAs to the quantum Yang-Baxter
equation.
\begin{theo}
If $(A, R)$ is a quasi-triangular Hopf algebra then $R$
satisfies the QYBE
\be
R_{12}R_{13}R_{23}=R_{23}R_{13}R_{12}
\ee
\end{theo}
The way to prove this theorem is to
calculate $(1 \tepro \Delta^{\prime})R$
in two ways. Write $R=R^{(1)}_{i} \tepro R^{(2)}_{i}$
, where summation over $i$ is understood, then
we find
\begin{eqnarray}
(1 \tepro \Delta^{\prime})(R) & = & R^{(1)}_i \tepro
\Delta^{\prime}(R^{(2)}_i) \nonumber \\
& = & \mbox{} R^{(1)}_i \tepro R \Delta (R^{(2)}_i)R^{-1}
\nonumber \\
& = & \mbox{} (1 \tepro R)(R^{(1)}_i \tepro \Delta (R^{(2)}_i
))(1 \tepro R^{-1}) \nonumber \\
& = & \mbox{} R_{23}(1 \tepro \Delta )R R_{23}^{-1} \nonumber \\
& = & \mbox{} R_{23}R_{13}R_{12}R_{23}^{-1}
\end{eqnarray}
On the other hand we have
\begin{eqnarray}
(1 \tepro \Delta^{\prime})R & = & (1 \tepro \tau)(1 \tepro
\Delta )R \nonumber \\
& = & \mbox{} (1 \tepro \tau )R_{13}R_{12} \nonumber \\
& = & \mbox{} R_{12}R_{13}
\end{eqnarray}
Equating these two results we get the required result.

In the applications of the Yang-Baxter equation to knot
theory and conformal field theory the $R$-matrix satisfies
another relation associated to the fact that the braid
of figure 4
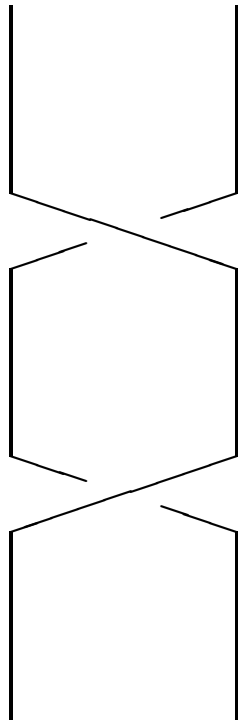
\begin{figure}
\centering
\setlength{\unitlength}{1mm}
\begin{picture}(50, 125)
\thicklines
\put(10, 110){\line(0, -1){25}}
\put(10, 85){\line(3, -1){30}}
\put(40, 75){\line(0, -1){25}}
\put(40, 50){\line(-3, -1){30}}
\put(10, 40){\line(0, -1){25}}
\put(40, 110){\line(0, -1){25}}
\put(40, 85){\line(-3, -1){10}}
\put(10, 75){\line(3, 1){10}}
\put(10, 75){\line(0, -1){25}}
\put(10, 50){\line(3, -1){10}}
\put(40, 40){\line(-3, 1){10}}
\put(40, 40){\line(0, -1){25}}
\end{picture}   \caption{This braid is equivalent to the trivial
braid}
\end{figure}
is equivalent to the trivial braid.
This relation reads $R_{12}R_{21}=1$.

\begin{definition}
A quasi-triangular Hopf algebra is called triangular if the
element $R$ satisfies the extra relation
\be
R_{12}R_{21}=1
\ee
\end{definition}

It is easy to think of an example of a QTHA albeit a rather
trivial one. Take for example the universal enveloping algebra
of any Lie algebra together with the element $R=1 \tepro 1$.
In fact this is a triangular Hopf algebra. The reason that
this example is trivial
is obviously the fact that universal enveloping algebras
are co-commutative. For ${\cal U}_q (sl_2)$ however $R=1 \tepro 1$
is not a quasitriangular structure since $\Delta ' \neq \Delta$
so the first axiom is violated. For $\hbar =0$ however it must again
be a solution which leads us to make the following ansatz for $R$
\be
R= 1 \tepro 1 + \sum_{n=1}^{\infty}R^{(n)} \hbar^{n}
\ee
Inserting this into the axioms 1 to 3 of a quasi-triangular
structure we get a set of recursive relations for the
elements $R^{n}$. Solving these relations leads to a solution
$R$ that satisfies all requirements. This method of constructing
$R$  is however extremely cumbersome for other cases than $sl_2$.
In the next section we therefore take a look at a systematic method
for constructing these $R$-matrices.

\subsection{The quantum double construction}
In this section we will show that the QUEA ${\cal U}_q (sl_2)$
is a quasi-triangular Hopf algebra. We will prove this by
explicitly constructing an element $R$ satisfying the axioms
of a quasi-triangular Hopf algebra. The construction we use
is the quantum version of the classical double construction
we considered earlier.

We start with some notational matters. Let
$A$ be a Hopf algebra and $A^{\ast}$ its dual Hopf
algebra. Replacing the comultiplication on $A^{\ast}$ by
the opposite comultiplication we obtain a new Hopf algebra
denoted by $A^{o}$. Choose in $A$ a basis $\{e_i\}$ and let
$\{e^i\}$ be the dual basis. Then we denote
\begin{eqnarray}
e_i .e_j & = & m(e_i \tepro e_j) = m^{k}_{ij}e_k \\
\Delta (e_i) & = & \Delta^{kl}_i e_k \tepro e_l \\
S(e_i) & = & S^{k}_{i}e_k
\end{eqnarray}
where we used the summation convention for repeated indices.
The relations in the algebra $A^o$ then become
\begin{eqnarray}
e^i . e^j & = & \Delta ^{ij}_{k}e^k \\
\Delta (e^i) & = & m^i_{lk} e^k \tepro e^l \\
S(e^i) & = & S^i_k e^k
\end{eqnarray}
where we used the definitions of the structure maps of the
dual Hopf algebra in terms of the structure maps of the
Hopf algebra itself.

Consider now the space $D(A)$ which, as a vectorspace, is isomorphic
to $A \tepro A^o$ and which contains $A$ and $A^o$ as Hopf
subalgebras. The structure of this space will become clearer in a
moment, however we can say the following. In the classical double
construction we considered the space $L \oplus L^{\ast}$ which
obviously contained $L$ and $L^{\ast}$ as Lie subalgebras. In the
quantum case however we are always working at the level of
universal enveloping algebras. At this level the classical double
construction gives a similar structure as the one we encounter here
because of the isomorphism
\be
{\cal U}(L_1 \oplus L_2) \simeq {\cal U}(L_1) \tepro
{\cal U}(L_2)
\ee
so the universal enveloping algebra of the double $D=L \oplus
L^{\ast}$ is isomorphic to \\ ${\cal U}(L) \tepro {\cal U}(L^{\ast})$
as a vectorspace and contains ${\cal U}(L)$ and ${\cal U}(L^{\ast})$
as subalgebras.

$D(A)$ is not yet an algebra itself because we do not know
how to multiply elements of $A$ with $A^o$ (we will come
to this later). Consider the element
$R=e_i \tepro e^i$ of $D(A) \tepro D(A)$. We have
\begin{eqnarray}
(\Delta \tepro 1)(R) & = & (\Delta \tepro 1)(e_i \tepro e^i)
\nonumber \\
& = & \mbox{} \Delta^{kl}_{i} e_k \tepro e_l \tepro e^i
\nonumber \\
& = & \mbox{} e_k \tepro e_l \tepro e^k.e^l \nonumber \\
& = & \mbox{} R_{13}R_{23}
\end{eqnarray}
In the same way we find for this $R$
\be
(1 \tepro \Delta)R = R_{13}R_{12}
\ee
so two of the three axioms for	a quasi-triangular Hopf
algebra are satisfied. The remaining axiom
$\Delta^{\prime}=R \Delta R^{-1}$ is easily seen to lead
to the equation
\be
\Delta^{lk}_i m^p_{kj} (e_l .e^j)=
\Delta^{kl}_i m^p_{jk} (e^j .e_l)
\ee
If we define the multiplication
between elements $e_i$ and $e^j$ such that
these permutation relations are satisfied, then the
element $R=e_i \tepro e^i$ (called the canonical element)
defines a quasi-triangular structure on $D(A)$.
The pair $(D(A), R)$ is then called the double of $A$.
Using the Hopf algebra axioms we can rewrite the
permutation relations into
\be
e^i .e_j=m^i_{kl}m^k_{nm}\Delta^{pl}_j\Delta^{sr}_p
S^n_s (e_r .e^m) \label{perm}
\ee
which is the form in which we shall use them later on.

Summarizing we have  the following theorem:
\begin{theo}
To every Hopf algebra $A$ there is associated a
quasi-triangular Hopf algebra that contains $A$
and $A^o$ as Hopf subalgebras and is isomorphic
to $A \tepro A^o$ as a vectorspace.
\end{theo}

We will now apply the above theory to find
an $R$ matrix making ${\cal U}_q(sl_2)$ into a quasi-triangular
Hopf algebra. Apart from the generators $(E, F, H)$ of
${\cal U}_q(sl_2)$ (the Chevalley generators) we will also
need the algebra in terms of generators $(e, f, H)$
where
\begin{eqnarray}
e & = & q^{H/2}E \\
f & = & q^{-H/2}F \\
H & = & H
\end{eqnarray}
The relations of ${\cal U}_q(sl_2)$ in terms of these
generators are
\begin{eqnarray}
[H, e] & = & 2e \\ {}
[H, f] & = & -2f \\ {}
[e, f] & = & \frac{2sinh(\frac{\hbar}{2}H)}{1-q^{-2}} \\
\Delta (e) & = & 1 \tepro e + e \tepro q^H \\
\Delta (f) & = & f \tepro 1 + q^{-H} \tepro f \\
\Delta (H) & = & H \tepro 1 + 1 \tepro H \\
S(e) & = & -q^{-H}e \\
S(f) & = & -q^{-2} q^{H}f \\
S(H) & = & -H
\end{eqnarray}

Considering the relations of ${\cal U}_q(sl_2)$ it is
not difficult to see that the sets
\be
\{H^n E^m F^l \}_{n, m, l=0}^{\infty}
\ee
and
\be
\{H^n e^m f^l \}_{n, m, l=0}^{\infty}
\ee
are bases of this algebra. Analogous to the case of
the classical UEA we define the positive and negative
Borel subalgebras
\begin{eqnarray}
{\cal U}_q (b_+) & = & span \{H^n E^m \}_{n, m=0}^{\infty} \\
{\cal U}_q (b_-) & = & span \{H^n F^m \}_{n, m=0}^{\infty}
\end{eqnarray}

What we will do now is construct the double
of the positive Borel subalgebra. Let $W$ and $Y$ be the
dual elements of $H$ and $E$ respectively, i.e.
\begin{eqnarray}
W(H) & = & 1 \\
W(E) & = & 0 \\
Y(H) & = & 0 \\
Y(E) & = & 1
\end{eqnarray}
Obviously $\{W^n Y^m \}_{n, m=0}^{\infty}$ is then a basis for
the dual of the positive Borel subalgebra,
$({\cal U}_q (b_+))^{\ast}$, dual to the basis
$\{H^n E^m\}$. Using the rule
\be
\langle f.g , x\rangle = \langle f \tepro g, \Delta (x) \rangle
\ee
we can calculate the commutation relation between
$W$ and $Y$. We find
\be
[W, Y]=-\frac{\hbar}{2}Y
\ee
In the same way, using the rule
\be
\langle \Delta (f), x \tepro y \rangle =
\langle f, xy \rangle
\ee
we can calculate the co-multiplication on $({\cal U}_q (b_+)
)^{\ast}$. The result of this easy exercise is
\begin{eqnarray}
\Delta (W) & = & W \tepro 1 + 1 \tepro W \\
\Delta (Y) & = & 1 \tepro Y + Y \tepro e^{-2W}
\end{eqnarray}
We see that if we define $\tilde{W}=\frac{4}{\hbar}W$ and
$\tilde{Y}=\frac{1-q^2}{\hbar}Y$
 and if we transpose the comultiplication
in order get the relations of the opposite dual
$({\cal U}_q (b_+))^o$, we get
\begin{eqnarray}
[\tilde{W}, \tilde{Y}] & = & -2 \tilde{Y} \\
\Delta (\tilde{W}) & = & \tilde{W} \tepro 1 + 1 \tepro \tilde{W} \\
\Delta (\tilde{Y}) & = & \tilde{Y} \tepro 1 + q^{-\tilde{W}}
\tepro \tilde{Y}
\end{eqnarray}
Compare these relations to the relations of the negative
Borel subalgebra. They are exactly equal if we take
$\tilde{W}=H$ and $\tilde{Y}=f$ which leads us to conclude
that
\be
({\cal U}_q (b_+))^o \cong {\cal U}_q (b_-)
\ee
The quantum double of the positive Borel subalgebra is
therefore the Hopf algebra ${\cal U}_q (b_+) \tepro {\cal U}_q(b_-)$
which is generated by the elements
$\{H, e, \hat{H}, f\}$ (where we denote the Cartan element
of the negative Borel subalgebra by $\hat{H}$). However,
we have not yet calculated the commutation relations between
elements of the positive Borel subalgebra and its opposite
dual.  Using eqn.(\ref{perm}) we can easily calculate these.
The resulting set of commutation relations for the double
algebra $D({\cal U}_q (b_+))$ is
\begin{eqnarray}
[H, e] & = & 2e \\ {}
[H, f] & = & -2f \\ {}
[\hat{H}, e] & = & 2e \\ {}
[\hat{H}, f] & = & -2f \\ {}
[e, f] & = & [H]_q \\ {}
[H, \hat{H}] & = & 0
\end{eqnarray}
 From these commutation relations it is immediate that
\be
\frac{D({\cal U}_q (b_+))}{\langle H-\hat{H} \rangle} \cong
{\cal U}_q (sl_2)
\ee
where $\langle H-\hat{H} \rangle$ is the ideal generated
by the element $H-\hat{H}$. The canonical homomorphism is
explicitly given by
\begin{eqnarray}
H, \hat{H} & \rightarrow & H \\
e & \rightarrow & e \\
f & \rightarrow & f
\end{eqnarray}

We can now easily construct the canonical element of the
double. We know that  $\{H^n e^m \}$ is a basis
of ${\cal U}_q (b_+)$ and $\{W^k Y^l\}$ is a basis of
$({\cal U}_q (b_+))^o$, however, these two bases are not
dual. Consider for a moment the general case again where
$\{e_i\}$ and $\{f^i\}$ are (not necessarily dual) bases
of $A$ and $A^o$ respectively. Also suppose that
$f^i =B^i_j e^j$ where $\{e^i\}$ is the dual basis of
$\{e_i\}$. Then the canonical element $R=e_i \tepro e^i$
w.r.t. the bases $e_i$ and $f^j$ is equal to
\be
R=(B^{-1})^i_j e_i \tepro f^j
\ee
and we also have $\langle f^i, e_j \rangle =B^i_j$.

We learn from this that we have to calculate the matrix
\be
B^{nm}_{kl} =\langle W^k Y^l, H^n e^m \rangle
\ee
and invert it. Explicit calculation shows that
\be
B^{nm}_{kl}=\delta^n_k \delta^m_l \frac{k![l;q^{-2}]!}{\hbar^l}
\ee
where we used the notation
\be
[u;q]!=\prod_{i=1}^{u}\frac{1-q^i}{1-q}
\ee
Fortunately the matrix $B$ is diagonal which makes inverting
it very easy. The result is
\be
(B^{-1})^{nm}_{kl}=\frac{\hbar^l}{k![l;q^{-2}]!} \delta^n_k
\delta^m_l
\ee
Finally we can write down the explicit expression for the
canonical element $R$:
\begin{eqnarray}
R & = & \sum_{kl} \frac{\hbar^l}{k![l;q^{-2}]!} H^k e^l \tepro
W^k Y^l \nonumber \\
& = & e^{H \tepro W} \sum_l \frac{\hbar^l}{[l;q^{-2}]!}
e^l \tepro Y^l
\end{eqnarray}
The image of this element under the canonical homomorphism
$D({\cal U}_q (b_+)) \rightarrow {\cal U}_q (sl_2)$ is therefore
\be
R=q^{\frac{1}{2}H \tepro H} E_{q^{-2}}^{\lambda e \tepro f}
\ee
where $\lambda =1-q^{-2}$ and
\be
E_q^x \equiv \sum_{l=0}^{\infty} \frac{x^l}{[l;q]!}
\ee
is the so called q-deformed exponential.

In the standard (fundamental) representation of $sl_2$,
which is also a representation of ${\cal U}_q (sl_2)$, this matrix
takes on the form
\be
R=
\left(
\begin{array}{cccc}
q & 0 & 0 & 0  \\
0 & 1 & 0 & 0  \\
0 & 1-q^{-2} & 1 & 0 \\
0 & 0 & 0 & q
\end{array}
\right)
\ee

This concludes the derivation of the quasi-triangular
structure of ${\cal U}_q (sl_2)$.

\subsection{The classical limit}
In this paper we have considered two Yang-Baxter equations, the
classical YBE and the quantum YBE. One might wonder what the
relation between them is. This is contained in the following
theorem.
\begin{theo}
Let $R(t)$ be a one parameter family of solutions of the
quantum YBE which can be written as a power series
of the form
\be \label{power}
R(t)= 1 \tepro 1 + r t + A t^2 + {\cal O}(t^3 )
\ee
then $r$ satisfies the classical YBE.
\end{theo}
Proof: Inserting the power series (\ref{power}) into the QYBE
\be
R_{12}R_{13}R_{23}=R_{23}R_{13}R_{12}
\ee
we get
\begin{eqnarray*}
\lefteqn{(r_{12}r_{13}+r_{12}r_{23}+r_{13}r_{23})
t^2 + {\cal O}(t^3)= } \\
& & (r_{13}r_{12}+r_{23}r_{12}+r_{23}r_{13})t^2 +
{\cal O}(t^3 )
\end{eqnarray*}
from which the theorem follows immediately.

The solution of the QYBE constructed above using the
quantum double construction
can be expanded in a power series of
the deformation parameter $\hbar$. The associated solution of the
classical YBE is easily found to be (up to overall rescaling
which is irrelevant because the YBEs are homogeneous)
\be
r=\frac{1}{2}H \tepro H + 2 E \tepro F
\ee
which is the $r$-matrix considered in example 3.

\section{Elements of representation theory}
In this last section we shall be taking a look at the
representation theory of ${\cal U}_q (sl_2)$. We will not
go into great detail however and will not prove all the
theorems (see for example \cite{PaSa,RoAr,ReSm,Kel,Ki,Re}).
What we will do is
derive the irreducible (physical) representions of the algebra
${\cal U}_q (sl_2)$ at $q$ a root of unity (i.e.
$q^m =1$ for some $m$), discuss the indecomposable
(non-physical) reps and also the tensorproduct decompositions
(fusion rules) of tensorproduct representations. The reason
why we consider the representation theory only at $q$ a root
of unity is that in the applications of quantum groups to
physics it is these representations that appear to be the most
important. The representation theory at $q$ not a root of unity
was given in \cite{Ros2} and is not too different from the
representation theory of ordinary simple Lie algebras. At a
root of unity this changes drastically. The presentation we
give here will follow closely the paper \cite{Kel}.

Let $q=e^{2 \pi i n/m}$ where $n, m$ are natural numbers,
$1 \leq n \leq m-1$ and $n, m$ are relatively prime. We set
$M=m$ for $m=$odd and $M=m/2$ for $M=$even. Consider
${\cal U}_q(sl_2)$ at these values of $q$
\begin{eqnarray}
[H, E] & = & E \;\;\;\;\;\;\; [H, F]=-F \;\;\;\;\;\;\; [E, F]=[2H]_q \\
\Delta (E) & = & E \tepro q^H + q^{-H} \tepro E \\
\Delta (F) & = & F \tepro q^H + q^{-H} \tepro F \\
\Delta (H) & = & H \tepro 1 + 1 \tepro H
\end{eqnarray}
(where we rescaled $H \rightarrow \frac{1}{2}H$) supplemented
by the relations
\be
E^M = F^M =0
\ee
which have to be imposed if the quantum
$R$-matrix is to still be defined for these values of $q$
(i.e. we insist on having a quasi-triangular structure because
we want the reps $V \tepro W$ and $W \tepro V$ to be
equivalent).

Now, let $\rho :{\cal U}_q (sl_2) \rightarrow End(W)$ be an
irreducible representation on a finite dimensional linear space $W$.
\begin{lemma}
$W$ is spanned by eigenvectors of $\rho (H)$.
\end{lemma}
Any matrix can be put in upper triangular form by a suitable
basis change in $W$. We put $\rho (H)$ in uppertriangular
form. Any upper triangular matrix has at least one eigenvector
$v$, namely $(1, 0, ...., 0)$. Using the commutation rules we easily
find
\be
\rho (H) (\rho (e^r f^s)v)=(\lambda +r-s)(\rho (e^r f^s)v)
\ee
 for $\rho (H) v=\lambda v$, so $\rho (e^r f^s)v$ are all
eigenvectors with different eigenvalues. Since $W$ is
irreducible a basis of $W$ must be contained in this set
of vectors (for $0 \leq r, s \leq M-1$). The Lemma follows.

Since dim$(W) < \infty$ there exists a highest weight vector
$\psi$ such that $\rho (e) \psi =0$ and $\rho (H) \psi =j \psi$
, where $j$ is the highest weight. It is easy to see that the set
\be
span \{\rho (f^r)\psi \mid 0 \leq r \leq M-1 \}
\ee
is invariant under $\rho ({\cal U}_q (sl_2)$ and is therefore
equal to $W$ (or else $W$ would not be irreducible). Also
note that dim$(W) \leq M$. From now on we denote $p=dim(W)$.
\begin{lemma}
\be
[ E, F^{k} ] = F^{k} [k]_{q} [2H-k+1]_{q}
\ee
\end{lemma}
The proof of this lemma is a straightforeward calculation
using the commutation relations of the algebra.

We now come to the main theorem.
\begin{theo}
The finite dimensional irreducible representations $W$ of
${\cal U}_q(sl_2)$ fall into two classes.
\begin{enumerate}
\item $dim(W)<M$: The inequivalent irreps are labeled by their
dimension $p$ and an integer $z$. They have highest weight
\be
j=\frac{1}{2}(p-1)+\frac{m}{4n}z
\ee
\item
$dim(W)=M$: The irreps are labeled by a complex number $z$
which ranges over the complex numbers minus the set
$\{ {\cal Z}+\frac{2n}{m}r \mid 1 \leq r \leq M-1\}$ and
have highest weight $j= \frac{1}{2}(M-1)+\frac{m}{4n}z$.
\end{enumerate}
\end{theo}
The irreducible representations  will be denoted by
$ \langle p, z \rangle $.

The proof of the theorem is not very difficult. Since the
dimension of $W$ is $p$ we have the following two
identities.
\begin{eqnarray}
\rho (EF^p) \psi & = & [p]_q[2j-p+1]_q \rho (F^{p-1})\psi =0
\label{een} \\
\rho (EF^r) \psi & = & [r]_q[2j-p+1]_q \rho (F^{r-1})\psi \neq 0
\end{eqnarray}
($r<p$) where we used the previous lemma. First consider the
case $p<M$. In that case $[p]_q \neq 0$ so the first equation
gives $[2j-p+1]_q=0$ which is easily seen to be equivalent to
\be
j =j(p, z) =\frac{1}{2}(p-1)+\frac{m}{4n}
\ee
where $z$ is an arbitrary integer. One can easily show that
within the parameters in which $p$ and $z$ are defined we have
an equality $j(p, z)=j(q, w)$ if and only if $p=q$ and $z=w$.
 From this it follows that $\rho (EF^r)\psi \neq 0$ for $r<p$
and also that $p$ and $r$ completely determine the irrep
up to isomorphism. For $p=M$ eqn (\ref{een}) is automatically
satisfied which means that the only demand on $j$ is
$[2j-r+1]_q \neq 0$ for $r \leq M-1$. From this it follows
that $j$ can take on any complex value except the values
$\{j(p, z)\mid 1 \leq p \leq M-1 ; z =integer\}$. Part 2 of
the theorem follows if we parametrize $j$ by $j=1/2(M-1)+
mz/4n$ where $z$ can be any complex number except for the
numbers specified in the theorem.

The fact that all irreps have dimension smaller or equal to $M$
does not mean that there are no representations with a higher
dimension. In fact if we consider the decomposition of a tensor
product representation of two irreducible reps we find not only
representations of the type discussed above but also some
indecomposable representations. By an indecomposable
representation we mean a representation which has invariant subspaces
but cannot be written as a direct sum of them. Without
proof we now give the decomposition of a tensor   product
rep of two irreducible representations.
\begin{theo}
\be
\langle i, z \rangle \tepro \langle j, w \rangle =
\bigoplus_{k=\mid i-j\mid+1}^{K} \langle k, z+w \rangle \oplus
\bigoplus_{l=r, r+2, ..}^{i+j-M} I^{l}_{z+w}
\ee
where $K=min\{i+j-1, 2M-i-j-1\}$.
\end{theo}
The representations $I^l_z$ are the indecomposable ones.
A few features of their structure are collected
in the next theorem.
\begin{theo}
The indecomposable representations have the following properties:
\begin{itemize}
\item $dim(I^l_z)=2M$
\item $I^l_z$ is a direct sum of weight spaces $W_{\mu}$
with highest weight
$\mu_{HW}(l, z)=\frac{1}{2}(M+p-2)+\frac{m}{4n}z$. The other weights
are $\{\mu_{HW}, \mu_{HW}-1, ..., \mu_{HW}-(M+p-1)\}$
\item $dim(W_{\mu})=2$ for the weights $\mu$ such that
\be
\frac{1}{2}(M-p)+(\frac{m}{4n}z) \geq \mu \geq -\frac{1}{2}(M-p)
+\frac{m}{4n}z
\ee
and $dim(W_{\mu})=1$ in all other cases.
\end{itemize}
\end{theo}
If one ignores these indecomposable
representations the tensor product
decompositions discribed above resemble
very much the fusion rules of certain
conformal field theories. Before you forget about them
however you must check if they form an ideal in the representation
ring. That this is indeed the case follows from the following
equations
\begin{eqnarray}
\langle j, z \rangle \tepro I^p_z &=& \bigoplus_{(l, n)}I^l_n \\
I^j_z \tepro I^p_w &=& \bigoplus_{(l, w)} I^l_n
\end{eqnarray}
The indecomposable representations can nicely be characterized
by the fact that their so called quantum dimensions are zero.
The quantum dimension $\chi^1_{\rho}$ of a representation $\rho$
is a special case of a character $\chi^r_{\rho}$ (where
$r$ is a complex number) defined as
\be
\chi^r_{\rho}=Tr(\rho(q^{2rH}))
\ee
We find for the representations discussed above
\begin{eqnarray}
\chi^1_{\langle p, z \rangle} & = & [p]_q e^{i \pi z} \neq 0 \\
\chi^1_{\langle p, 0 \rangle} & = & >0 \\
\chi^1_{I^j_w} & = & 0
\end{eqnarray}
so that the physical representations always have a positive
quantum dimension while the non-physical reps have zero
quantum dimension.

\section{Acknowledgements}
I would like to thank F.~A.~Bais, H.~W.~Capel and N.~D.~Hari~Dass
for reading the manuscipt and making useful comments.
I am also grateful for the help of A.~Hulsebos and R.~Rietman for
getting the layout of the paper as it is.


\begin{thebibliography}{99}
\bibitem{KuRe} P.\ P.\ Kulish, N.\ Yu.\ Reshetikhin {\em Quantum linear
problem for the sine-Gordon equation and higher representations},
 J.\ Sov.\ Math 23  (1983)  2435-2441
\bibitem{Skl} E.\ K.\ Sklyanin {\em Some algebraic structures connected
with the Yang-Baxter equation} Funct.\ Anal.\ Appl.\ 16  (1982)  263 and
Funct.\ Anal.\ Appl.\ 17 (1983) 273
\bibitem{Dr1} V.\ G.\ Drinfeld {\em Quantum groups} Proc.\  of the
international congress of mathematicians, Berkeley, 1986,
American Mathematical Society, 1987, 798
\bibitem{Dr2} V.\ G.\ Drinfeld {\em Hopf algebras and the quantum
Yang-Baxter equation} Sov.\ Math.\ Dokl.\ 32 (1985) 254
\bibitem{Dr3} V.\ G.\ Drinfeld {\em A new realization of Yangians and
quantized affine algebras} Sov.\ Math.\ Dokl.\ 36 (1988) 212
\bibitem{Dr4} V.\ G.\ Drinfeld {\em Hamiltonian structures on Lie
groups, Lie bi-algebras and the geometrical meaning of the
classical Yang-Baxter equations} Sov.\ Math.\ Dokl.\ 27 (1983) 68
\bibitem{Jim1} M.\ Jimbo {\em A q-difference analogue of
${\cal U} (g) $ and the Yang-Baxter equation} Lett.\ Math.\ Phys.\  10
 (1985) 63
\bibitem{Jim2} M.\ Jimbo {\em A q-analogue of ${\cal U} (gl (N+1) ) $,
Hecke algebra and the Yang-Baxter equation} Lett.\ Math.\ Phys.\
10 (1986) 247
\bibitem{Conf} L.\ Alvarez-Gaum\'e, C.\ Gomez, G.\ Sierra {\em Hidden
quantum symmetry in rational conformal field theories} Nucl.\ Phys.\
B310 (1989) ; {\em Quantum group interpretation of some conformal
field theories} Phys.\ Lett.\ 220B (1989) 142; {\em Duality and quantum
groups} Nucl.\ Phys.\ B330 (1990) 347\\
P.\ Furlan, A.\ Ch.\ Ganchev, V.\ Petkova
{\em Quantum groups and fusion multiplicities} Preprint INFN/AE-89/15,
Trieste 1989\\
G.\ Moore, N.\ Yu.\ Reshetikhin {\em A comment on
quantum symmetry  in conformal field theory} Nucl.\ Phys.\ B328 (1989)
557 \\ N.\ Yu.\ Reshetikhin, F.\ Smirnov {\em Hidden quantum group
symmetry and integrable perturbations of conformal field theory}
 Comm.\ Math.\ Phys.\  131 (1990) 157  \\
G.\ Mack, V.\ Schomerus {\em conformal algebras with quantum symmetry
from the theory of superselection sectors} Comm.\ Math.\ Phys.\ 134 (1990)
139 \\
G.\ Mack, V.\ Schomerus {\em Quasi Hopf quantum symmetry in quantum
theory} Preprint DESI 91-037 ISSN 0418-9833  (1991) \\
J. \ Fuchs, P. \ van Driel {\em WZW fusion rules, quantum groups
, and the modular matrix S} Nucl. \ Phys.\ B346 (1990) 632
\bibitem{Wor1} S.\ L.\ Woronowicz {\em Compact matrix pseudogroups}
Comm.\ Math.\ Phys.\ 111 (1987) 613
\bibitem{Wor2} S.\ L.\ Woronowicz {\em Tannaka-Krein duality for
compact matrix pseudogroups.\  Twisted $SU (N) $ groups} Invent.\ Math.\
93 (1988) 35
\bibitem{Wor3} S.\ L.\ Woronowicz {\em Differential calculus on
compact matrix pseudogroups  (quantum groups)  } Comm.\ Math Phys.\
122 (1989) 125
\bibitem{Man} Y.\ I.\ Manin {\em Quantum groups and non-commutative
geometry} Preprint Montreal Univ.\  CRM-1561  (1988)
\bibitem{Abe} E.\ Abe {\em Hopf algebras} Univ.\ Press, Cambridge 1980
\bibitem{Mil} J.\ W.\ Milnor, J.\ C.\ Moore {\em On the structure of Hopf
algebras} Ann.\ Math.\ 81 (1965) 211
\bibitem{sem} M.\ A.\ Semenov-Tian-Shansky {\em What is a classical
r-matrix} Funct.\ Anal. \ Appl. \ 17 (1983) 259
\bibitem{BFFLS} F.\ Bayen, M.\ Flato, C.\ Fronsdal, A.\ Lichnerowicz
and D.\ Sternheimer {\em Deformation theory and quantization} Ann.\
Phys.\ 111 (1978) 61-151
\bibitem{Ros1} M.\ Rosso {\em An analogue of P.\ B.\ W.\  theorem and the
universal $R$ matrix for ${\cal U} (sl_{N+1}) $} Comm.\ Math.\ Phys.\
124 (1989) 307
\bibitem{Ros2} M.\ Rosso {\em Finite dimensional representations of
the quantum analogue of the universal enveloping algebra of a
complex simple Lie algebra} Comm.\ Math.\ Phys.\ 117 (1988) 581
\bibitem{Bur1} N.\ Burroughs {\em Relating the approaches to
quantized algebras and quantum groups} preprint DAMTP/R-89/11, 1989
\bibitem{Bur2} N.\ Burroughs {\em The universal $R$-matrix for
${\cal U} (sl (3) ) $ and beyond} Comm.\ Math.\ Phys.\ 127 (1990) 109
\bibitem{Majid} S.\ Majid {\em Quasitriangular Hopf algebras and
Yang-Baxter equations} Int.\ J.\ Mod.\ Phys.\ A, vol5, 1 (1990) 1
\bibitem{PaSa} V.\ Pasquier, H.\ Saleur, Nucl.\ Phys.\ B330 (1990) 523
\bibitem{RoAr} P.\ Roche, D.\ Ardaudon, Lett.\ Math.\ Phys.\ 17 (1989) 295
\bibitem{ReSm} N.\ Reshetikhin, F.\ Smirnov, Comm.\ Math.\ Phys.\ 131 (1990)
157
\bibitem{Kel} G.\ Keller {\em Fusion rules of ${\cal U} (sl_2) $, $q^m=1$}
Lett.\ Math.\ Phys.\ 21 (1991) 273
\bibitem{Ki} A.\ N.\ Kirillov, N.\ Yu.\ Reshetikhin {\em Representations
of the algebra ${\cal U}_q (sl_2) $, $q$-orthogonal polynomials
and invariants of links} LOMI Preprint E-9-88 Leningrad 1988
\bibitem{Re} N.\ Yu.\ Reshetikhin {\em Quantized universal
enveloping algebras, the Yang-Baxter equation and invariants
of links} LOMI Preprint E-4-87, Leningrad 1988
\end{thebibliography}
\end{document}